\documentclass[utf8]{frontiersSCNS}

\usepackage{lineno,microtype,subcaption}
\usepackage[onehalfspacing]{setspace}
\usepackage{booktabs}

\usepackage[bookmarks=false,  pdfnewwindow=true, colorlinks=true,  linkcolor=magenta, citecolor=blue, filecolor=cyan, urlcolor=purple, final=true]{hyperref}

\def\keyFont{\fontsize{8}{11}\helveticabold }

\def\firstAuthorLast{Palmerio {et~al.}}
\def\Authors{Erika~Palmerio\,$^{1,2,*}$, Nariaki~V.~Nitta\,$^{3}$, Tamitha~Mulligan$^{4}$, Marilena~Mierla\,$^{5,6}$, Jennifer~O’Kane\,$^{7}$, Ian~G.~Richardson\,$^{8,9}$, Suvadip~Sinha\,$^{10}$, Nandita~Srivastava\,$^{11,10}$, Stephanie~L.~Yardley\,$^{7}$, and Andrei~N.~Zhukov\,$^{5,12}$}

\def\Address{\small{$^{1}$Space Sciences Laboratory, University of California--Berkeley, Berkeley, CA, USA\\
$^{2}$CPAESS, University Corporation for Atmospheric Research, Boulder, CO, USA\\
$^{3}$Lockheed Martin Solar and Astrophysics Laboratory, Palo Alto, CA, USA\\
$^{4}$Space Sciences Department, The Aerospace Corporation, Los Angeles, CA, USA\\
$^{5}$Solar--Terrestrial Centre of Excellence, Royal Observatory of Belgium, Brussels, Belgium\\
$^{6}$Institute of Geodynamics of the Romanian Academy, Bucharest, Romania\\
$^{7}$Mullard Space Science Laboratory, University College London, Dorking, UK\\
$^{8}$Department of Astronomy, University of Maryland, College Park, MD, USA\\
$^{9}$Heliospheric Physics Division, NASA Goddard Space Flight Center, Greenbelt, MD, USA\\
$^{10}$CESSI, Indian Institute of Science Education and Research Kolkata, Mohanpur, India\\
$^{11}$Udaipur Solar Observatory, Physical Research Laboratory, Udaipur, India\\
$^{12}$Skobeltsyn Institute of Nuclear Physics, Moscow State University, Moscow, Russia}}

\def\corrAuthor{Erika Palmerio}
\def\corrEmail{epalmerio@berkeley.edu}

\begin{document}
\onecolumn
\firstpage{1}

\title[Remote-sensing Techniques for Stealth CMEs]{Investigating Remote-sensing Techniques \\ to Reveal Stealth Coronal Mass Ejections} 

\author[\firstAuthorLast ]{\Authors} 
\address{} 
\correspondance{} 
\extraAuth{}

\maketitle


\begin{abstract}

\section{}
Eruptions of coronal mass ejections (CMEs) from the Sun are usually associated with a number of signatures that can be identified in solar disc imagery. However, there are cases in which a CME that is well observed in coronagraph data is missing a clear low-coronal counterpart. These events have received attention during recent years, mainly as a result of the increased availability of multi-point observations, and are now known as `stealth CMEs'. In this work, we analyse examples of stealth CMEs featuring various levels of ambiguity. All the selected case studies produced a large-scale CME detected by coronagraphs and were observed from at least one secondary viewpoint, enabling \emph{a priori} knowledge of their approximate source region. To each event, we apply several image processing and geometric techniques with the aim to evaluate whether such methods can provide additional information compared to the study of ``normal'' intensity images. We are able to identify at least weak eruptive signatures for all events upon careful investigation of remote-sensing data, noting that differently processed images may be needed to properly interpret and analyse elusive observations. We also find that the effectiveness of geometric techniques strongly depends on the CME propagation direction with respect to the observers and the relative spacecraft separation. Being able to observe and therefore forecast stealth CMEs is of great importance in the context of space weather, since such events are occasionally the solar counterparts of so-called `problem geomagnetic storms'.

\tiny
 \keyFont{ \section{Keywords:} Sun, coronal mass ejection (CME), stealth CME, solar corona, space weather, remote-sensing observations}
\end{abstract}


\section{Introduction}

Coronal mass ejections (CMEs) are powerful solar eruptions containing large amounts of plasma and magnetic field that are regularly expelled from the Sun into the heliosphere. They were first identified in white light in the early 1970's \citep[][]{tousey1973,gosling1974} in images from the $7^{\text{th}}$ Orbiting Solar Observatory (OSO-7) coronagraph \citep{koomen1975} and the coronagraph onboard the Skylab space station that formed part of the Apollo Telescope Mount \citep[ATM;][]{tousey1977} suite of solar instruments. Around the same time, the low-coronal counterparts of white-light CMEs were being observed through multi-wavelength solar disc imagery, e.g. in extreme ultra-violet (EUV), soft X-rays, and H$\alpha$, using data from OSO-7, Skylab, and ground-based observatories \citep[e.g.,][]{demastus1973,munro1979}. The typical low-coronal signatures of CMEs were reviewed by \citet{hudson2001}. These include the appearance of coronal dimmings \citep[e.g.,][]{thompson2000, kahler2001}, flare ribbons \citep[e.g.,][]{rust1973,martin1979}, post-eruption arcades \citep[e.g.,][]{rust1977,tripathi2004}, and EUV waves \citep[e.g.,][]{thompson1998,zhukov2004}, as well as the disappearance of filament material \citep[e.g.,][]{rust1975,sheeley1975} and X-ray sigmoids \citep[e.g.,][]{rust1996,green2007}. A major step forward in CME research was achieved in the early 1980's, when \citet{howard1982} reported observations of the first Earth-directed CME (in that case, a halo CME\footnote{Note that a halo CME is not necessarily Earth-directed, and a CME does not need to be a halo in order to impact Earth.}) in white-light images from the Solwind coronagraph \citep{michels1980} and linked this `coronal transient'\footnote{CMEs have been called by various names since their discovery, ranging from `coronal transient phenomena' \citep{macqueen1974} to `mass ejections from the Sun' \citep{gosling1974}. The term `coronal mass ejection' was first introduced by \citet{gosling1975}, but came into common usage only in the early 1980's.} with a disappearing filament on the solar disc and with a shock wave detected near Earth about 3 days later. The following decade saw the launch of two missions that have made a major impact on the field of solar physics, namely Yohkoh \citep[also known as Solar-A;][]{ogawara1991} and the Solar and Heliospheric Observatory \citep[SOHO;][]{domingo1995}. Yohkoh carried several solar imagers, including the Soft X-ray Telescope \citep[SXT;][]{tsuneta1991}, whilst the SOHO payload includes remote-sensing and in-situ instruments, several of which are still operational, including the venerable Large Angle Spectroscopic Coronagraph \citep[LASCO;][]{brueckner1995}. By then, the whole Sun-to-Earth picture of CMEs was seemingly quite clear: The presence of low-coronal signatures preceding a halo CME observed by LASCO would signify that the CME is Earth-directed, whereas the lack of visible activity on the solar disc would indicate that the CME was associated with a far-sided eruption.

However, it was not long before this picture was shown to not always hold true. Studies of CMEs detected in situ near Earth noted that a large number of such events lacked clear solar associations \citep[e.g.,][]{cane2003,richardson2010}. Specifically, \citet{schwenn2005} reported that 20\% of interplanetary CMEs could not be linked to a front-sided (partial or full) halo CME source, and \citet{zhang2007} reported that 11\% of CME-driven storms with minimum $Dst\le -100$~nT could not be linked to eruptive signatures in the low corona. A turning point came with the launch of the Solar Terrestrial Relations Observatory \citep[STEREO;][]{kaiser2008} mission in 2006. The STEREO mission consisted of twin spacecraft, one advancing ahead of Earth in its orbit (STEREO-A) and one trailing behind (STEREO-B), thus enabling observations of the Sun from multiple viewpoints. Using data from the Sun Earth Connection Coronal and Heliospheric Investigation \citep[SECCHI;][]{howard2008a} suite onboard both STEREO spacecraft when they were separated by $53^{\circ}$, \citet{robbrecht2009} reported that a CME erupted \emph{``leaving no trace behind on the solar disc''}. Although the authors were able to identify a slow, streamer-blowout CME erupting off the limb in STEREO-A imagery, there were no corresponding on-disc signatures in STEREO-B data. The \citet{robbrecht2009} event represents the first direct observation of what is now known in the solar physics community as a `stealth CME'. In the following years, numerous additional stealth events have been reported \citep{ma2010,dhuys2014,kilpua2014}. According to these studies, CMEs that lack distinct low-coronal signatures tend to occur close to solar minimum, are generally slow and narrow, and often form at higher altitudes in the solar atmosphere. One major question that started to be raised is whether the eruption mechanism for stealth CMEs is fundamentally different from that of ``ordinary'' CMEs, or if stealth CMEs simply represent the lowest end of the full energy spectrum of solar eruptions \citep{howard2013,lynch2016}. Additionally, it was soon clear that stealth CMEs can be detected in situ, starting with the \citet{robbrecht2009} event that was observed at STEREO-B featuring a classic flux-rope structure \citep{mostl2009,lynch2010,nieveschinchilla2011}. It also became evident that the interplanetary counterparts of stealth CMEs are able to cause significant space weather disturbances if they encounter Earth \citep{nitta2017}. CME-driven storms that cannot be linked to a clear source on the Sun or to appreciable solar activity are particularly challenging to forecast and are known as `problem geomagnetic storms' \citep[e.g.,][]{mcallister1996}.

During the last decade, routine observations of the solar disc have advanced considerably due to the high temporal and spatial resolution of data from the Solar Dynamics Observatory \citep[SDO;][]{pesnell2012}, launched in 2010. Nevertheless, despite the improvement of EUV observations with respect to previous instrumentation, stealth CMEs continue to be reported \citep[e.g.,][]{nitta2017}. It has been suggested that apparently stealth CMEs result from observational limitations such as instrument sensitivity and bandwidth issues, even in the SDO era \citep{howard2013}, and that advanced image processing techniques may reveal hard-to-observe signatures in both solar disc and coronagraph imagery \citep{alzate2017,okane2019}. Additionally, some studies have applied geometric triangulation and reconstruction techniques to data from complementary viewpoints in order to trace stealth CMEs back to an approximate source region on the disc \citep{pevtsov2012,okane2019,talpeanu2020}. These methods, however, have not been tested on a large number of events and hence it is not known whether they are suitable to all circumstances or whether they can be applied only to a limited number of cases. To complicate things further, there is currently no formal definition describing what a stealth CME is or defining the ``observational limit'' below which a CME can be considered to be stealthy.

In this paper, we contemplate the following question: Since stealth CMEs can present diverse characteristics, is it possible that there exists a class of ``extremely stealth'' CMEs that cannot be revealed even with the aid of state-of-the-art techniques? In order to address this issue, this work aims to investigate the efficacy of various techniques applied to remote-sensing data in revealing the signatures of CMEs that are elusive on the solar disc. We test such techniques on four well-studied stealth CMEs for which the ``true'' source is more or less known because of the availability of remote-sensing imagery from additional viewpoints. This will ensure that any solar activity that is observed away from the expected source region will not be mistakenly interpreted as a signature of the stealth CME under analysis. This manuscript is organised as follows. In Section~\ref{sec:techniques}, we present and describe the imaging and geometric techniques that we employ in this study to analyse stealth CMEs. In Section~\ref{sec:analysis}, we apply these techniques to four case studies and compare them with information that can be retrieved from plain inspection of intensity images only. Finally, in Section~\ref{sec:conclusions} we discuss our results and present our conclusions.


\section{Techniques Employed} \label{sec:techniques}

This section summarises the remote-sensing techniques that are used throughout this work to analyse the four elusive CMEs under study. Imaging techniques are described in Section~\ref{subsec:imaging}, whilst geometric techniques are presented in Section~\ref{subsec:geometric}.

\subsection{Imaging Techniques} \label{subsec:imaging}

\subsubsection{Image Differencing} \label{subsubsec:difference}

The image differencing technique simply consists of subtracting from an image a preceding one, so that changes in total intensity over time appear as patches that are either dark (denoting an intensity decrease) or bright (denoting an intensity increase). This method has been long used in solar physics applications for both on-disc and coronagraph observations \citep[e.g.,][]{burlaga1982,hudson1992}. The commonly used nomenclature for the technique is that subtracting a pre-event image yields a `base difference' image, whilst subtracting successive images yields a `running difference' image. Base-difference images are often used to highlight transient phenomena that develop over larger time scales, such as coronal dimmings \citep[e.g.,][]{attrill2010}, whilst running-difference images are often used to highlight short-term transient features such as EUV waves \citep[e.g.,][]{attrill2007}. Since stealth CMEs tend to be slow, i.e.\ they erupt and accelerate over the course of several hours, their evolution is not expected to be captured in running-difference images, hence we focus in this work on using image differencing over longer time scales. In particular, \citet{nitta2017} noted that in the case of stealth CMEs, difference images with ``long enough'' temporal separations (of the order of ${\sim}10$~hours) should often be used in order to reveal weak low-coronal signatures, including dimmings and post-eruption arcades. In this work, in order to minimise the appearance of artefacts at the solar limb due to rotation (which are still present even after accounting for differential rotation), we calculate the percentage variation between each couple of images and use a fixed temporal separation of ${\Delta}t = 12$~hours.

\subsubsection{Wavelet Packets Equalisation} \label{subsubsec:wpe}

The Wavelet Packets Equalisation \citep[WPE;][]{stenborg2003,stenborg2008} technique is a multi-resolution image processing method that can be applied to enhance features based on their multi-scale nature. In the WPE technique, an image is first decomposed over both dimensions using spatially localised functions known as wavelets. We have implemented the \citet{stenborg2003} procedure using a 2D \textit{\`{a} trous} wavelet transform \citep[e.g.,][]{shensa1992}, where the scaling function is a $B_3$-spline corresponding to a $5{\times}5$ smoothing kernel. This produces a set of wavelet planes at different spatial scales (called wavelet scales) derived from the initial image, together with the remaining ``continuum’’ background image representing the lowest frequencies and largest scales. A wavelet-processed image is constructed by summing over all the wavelet planes and the continuum background with user-defined weights to emphasise the desired scale sizes. In general, the weighting strategy must be fine-tuned to both the particular image (e.g., different EUV wavelengths or different coronagraphs) and to the spatial scales of interest. In the case of stealth CMEs, the WPE technique has been used to enhance off-limb structures from secondary viewpoints \citep[e.g.,][]{vourlidas2011,nieveschinchilla2013,liewer2021}, and had been explored to search for elusive on-disc signatures by \citet{robbrecht2009}.

\subsubsection{Multi-scale Gaussian Normalisation} \label{subsubsec:mgn}

The Multi-scale Gaussian Normalisation \citep[MGN;][]{morgan2014} technique, similarly to the WPE method described in Section~\ref{subsubsec:wpe}, is based on a multi-scale normalisation algorithm. In the MGN technique, a set of 2D Gaussian kernels of different scale lengths are used to locally normalise an image using a set of local mean and standard deviation values. A weighted combination of the normalised components is then used to obtain a weighted mean locally normalised image, which is finally superposed to the corresponding global gamma-transformed image. This results in an enhancement of the local intensity fluctuations within the images. The MGN image processing tool has been applied to EUV images to identify the low-coronal signatures associated with stealth events in both on-disc and off-limb observations \citep{alzate2017,okane2019,okane2021a}.

\subsection{Geometric Techniques} \label{subsec:geometric}

\subsubsection{Latitude Projection} \label{subsubsec:latproj}

Since all CMEs investigated in this study have been observed off limb from at least one additional perspective, a simple approach to adopt when triangulation is not possible (i.e., when only one ``non-stealthy'' viewpoint is available) is to project the approximate latitude from which the CME originated onto the ``stealthy'' field of view. Considering a ``classic'' three-part CME structure in coronagraph imagery \citep{illing1985}, we trace back to the solar disc the approximate latitude of the central, bright core that is observed off limb. We note that not all CMEs feature a three-part structure \citep[e.g.,][]{vourlidas2013}, but stealth CMEs often belong to the streamer blowout category, in which flux rope signatures tend to occur at a higher rate than in the general CME population \citep{vourlidas2018}. The projected latitude of the source region from the off-limb viewpoint naturally focuses the search for any possible faint or ambiguous on-disc signature to a more localised area. This enhances the potential to find any low-coronal signatures or dynamics that, by themselves, would not have necessarily been interpreted as being eruption-related.

\subsubsection{Tie-point Technique} \label{subsubsec:tp}

The Tie-point (TP) triangulation technique was formulated by \citet{inhester2006} and first employed by \citet{thompson2009} to study a Sun-grazing comet. The principle on which the TP technique is based is that two separate observers and the point of interest in space (to be triangulated) form a plane called `epipolar plane', which is reduced to a line (`epipolar line') in image projections. A point identified in an image from the first observer must lie on the same epipolar line in the corresponding image from the second observer. Larger-scale features can be tracked by finding correspondences between different pixels along epipolar line pairs in images from both spacecraft. The 3D reconstruction or triangulation is then achieved by finding the intersection of the two lines of sight (for each pixel of interest) along the corresponding epipolar plane, which is unambiguously defined. In solar physics applications, the TP method has been used to evaluate the 3D morphology of erupting filaments \citep[e.g.,][]{bemporad2011,thompson2012,panasenco2013,palmerio2021} and the evolution of CME fronts or cores in coronagraph data \citep[e.g.,][]{mierla2008,mierla2009,srivastava2009,liewer2011}.

\subsubsection{Graduated Cylindrical Shell} \label{subsubsec:gcs}

The Graduated Cylindrical Shell \citep[GCS;][]{thernisien2006,thernisien2009,thernisien2011} model is a reconstruction technique usually applied to white-light coronagraph images. In the GCS model, a wireframe describing the geometry of flux ropes is used to fit CMEs from one or more simultaneous viewpoints. Within a single reconstruction, six free parameters (latitude, longitude, axis tilt, apex height, half-angle, and aspect ratio) can be adjusted until they best match the CME morphology observed in one or more images. The geometry of the model itself is often referred to as a ``hollow croissant'' and consists of a half-torus frontal part with two conical legs connected to the Sun. The resulting shape, reminiscent of a croissant, is ``hollow'' in the sense that the electron density is placed uniquely on the shell of the model. Thus, fits performed with the GCS model can provide information on the morphology of CMEs, but not on their magnetic field structure. The GCS technique is widely used in solar physics and space weather applications to determine geometric and kinematic parameters of CMEs and their shocks through the corona \citep[e.g.,][]{mierla2010,shi2015,schmidt2016}, also in the case of stealth CMEs \citep[e.g.,][]{lynch2010,he2018}. More recently, \citet{okane2019} and \citet{freiherrvonforstner2021} used the inferred propagation latitude and longitude from GCS reconstructions to obtain an approximate location for the source region of the stealth CMEs they analysed. The full list of GCS-reconstructed parameters for all CMEs studied in this work can be found in Supplementary Table~1.


\section{Analysis of Sample Stealth Events} \label{sec:analysis}

We present in this section the remote-sensing analysis of four CMEs with elusive on-disc signatures: 2008 June 1, 2011 March 3, 2012 February 4, and 2016 October 8. These events were selected based on two main factors. First of all, they were all observed as classic three-part CMEs including a flux rope by at least one spacecraft, i.e., they do not belong to the so-called `jet' and/or `blob' categories. Furthermore, each event was observed from at least one additional viewpoint, enabling estimation of its approximate source region on the solar disc. Such observations are provided for each case study as supplementary videos in which EUV data have been enhanced with a radial filter, in order to bring out off-limb emission. The reader is invited to initially rapidly move the video player slider back and forth, so that the motion catches the eye. As stealth CMEs are usually slower than average, it may be difficult to identify erupting structures that evolve over extremely long time scales when played at the speeds shown in the videos (i.e., 2--4~hours per second). Additionally, kinematic (height--time) plots for each event based on observations from these additional viewpoints are provided in Figure~S1.

\subsection{Event 1: 2008 June 1} \label{subsec:2008jun}

The first CME that we focus on in this study (Event 1) erupted on 2008~June~1 and was a stealth event as seen from STEREO-B. This CME was first reported by \citet{robbrecht2009} and its eruption was later modelled by \citet{lynch2016}. As mentioned in the Introduction, this event marked the first direct observation of a CME that left \emph{``no trace behind''} in EUV imagery from one viewpoint, hence we treat it here as a ``stealth CME prototype''. At the time of this event, STEREO-A was located $28^{\circ}$ west of Earth and STEREO-B was positioned $25^{\circ}$ east of Earth. The pre-eruptive configuration and eruption process were well observed by STEREO-A, as shown in Extreme UltraViolet Imager (EUVI) and COR1 coronagraph data shown in Supplementary Video 1. From the STEREO-A perspective, a flux rope structure \citep[observed as a characteristic cavity in off-limb imagery;][]{gibson2006} can be seen to lie at relatively high altitudes below a coronal streamer above the southeastern limb. The eruption itself took place over large time scales (${\sim}$1.5~days), during which the flux rope slowly lifted off (starting around 15:00~UT on May~31), causing the streamer to swell and resulting in a classic streamer-blowout CME that reached the COR2-A coronagraph field of view around 22:00~UT on June~1. We remark that corresponding images from the Extreme-ultraviolet Imaging Telescope \citep[EIT;][]{delaboudiniere1995} onboard SOHO are not available because of a data gap.

\begin{figure}[!ht]
\begin{center}
\includegraphics[width=0.99\linewidth]{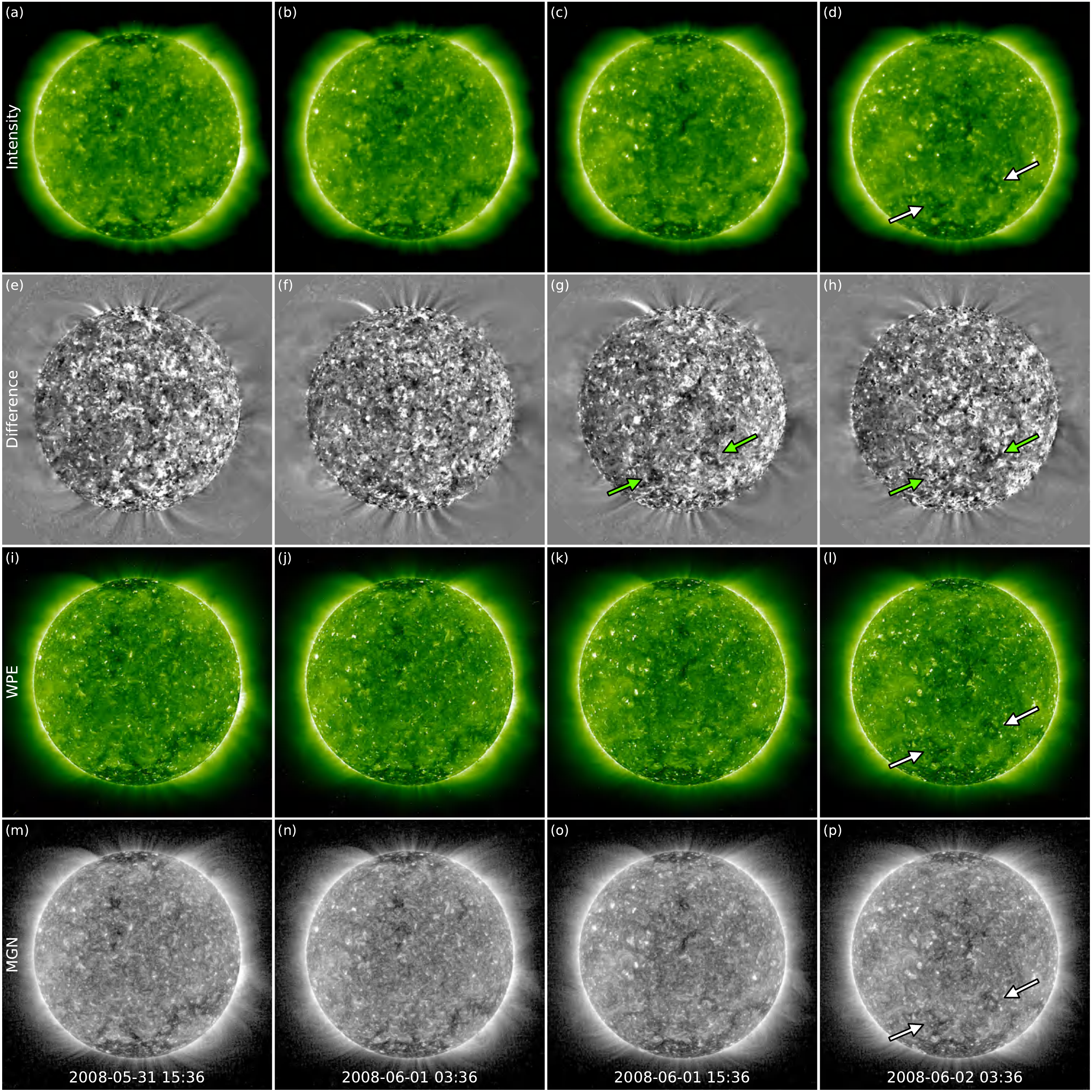}
\end{center}
\caption{Imaging techniques applied to the 2008 June 1 CME (Event 1). STEREO/SECCHI/EUVI-B 195~{\AA} images are shown at four different times and processed with four different methods. (a--d) Intensity images. (e--h) Difference images with fixed $\Delta t = 12$~hours. (i--l) WPE-processed images. (m--p) MGN-processed images. Two faint dimmings are indicated with arrows (see text for details).}
\label{fig:20080601_imaging}
\end{figure}

EUVI images from STEREO-B for the eruption period processed with different techniques are shown in Figure~\ref{fig:20080601_imaging} and Supplementary Video 2. It is clear that the succession of images in Figure~\ref{fig:20080601_imaging} does not show any strong indication that an eruption has occurred. Nevertheless, it is possible to note two extremely faint dimmings (indicated with arrows) developing in the southern hemisphere starting around 15:00~UT on June~1 (see also Supplementary Video~2). These dimmings are not straightforward to identify even in difference data, possibly because they are rather weak and hence appear ``camouflaged'' by other intensity fluctuations on the solar disc. We also note that these signatures appear equally visible in images produced using the other three methods (i.e., intensity, WPE, and MGN). The dimmings seem to be spatially consistent with the approximate CME source region deduced from STEREO-A imagery (see Supplementary Video 1), but because they are so faint it is not possible to draw strong conclusions as to their association with the 2008 June~1 CME.

\begin{figure}[!ht]
\begin{center}
\includegraphics[width=0.75\linewidth]{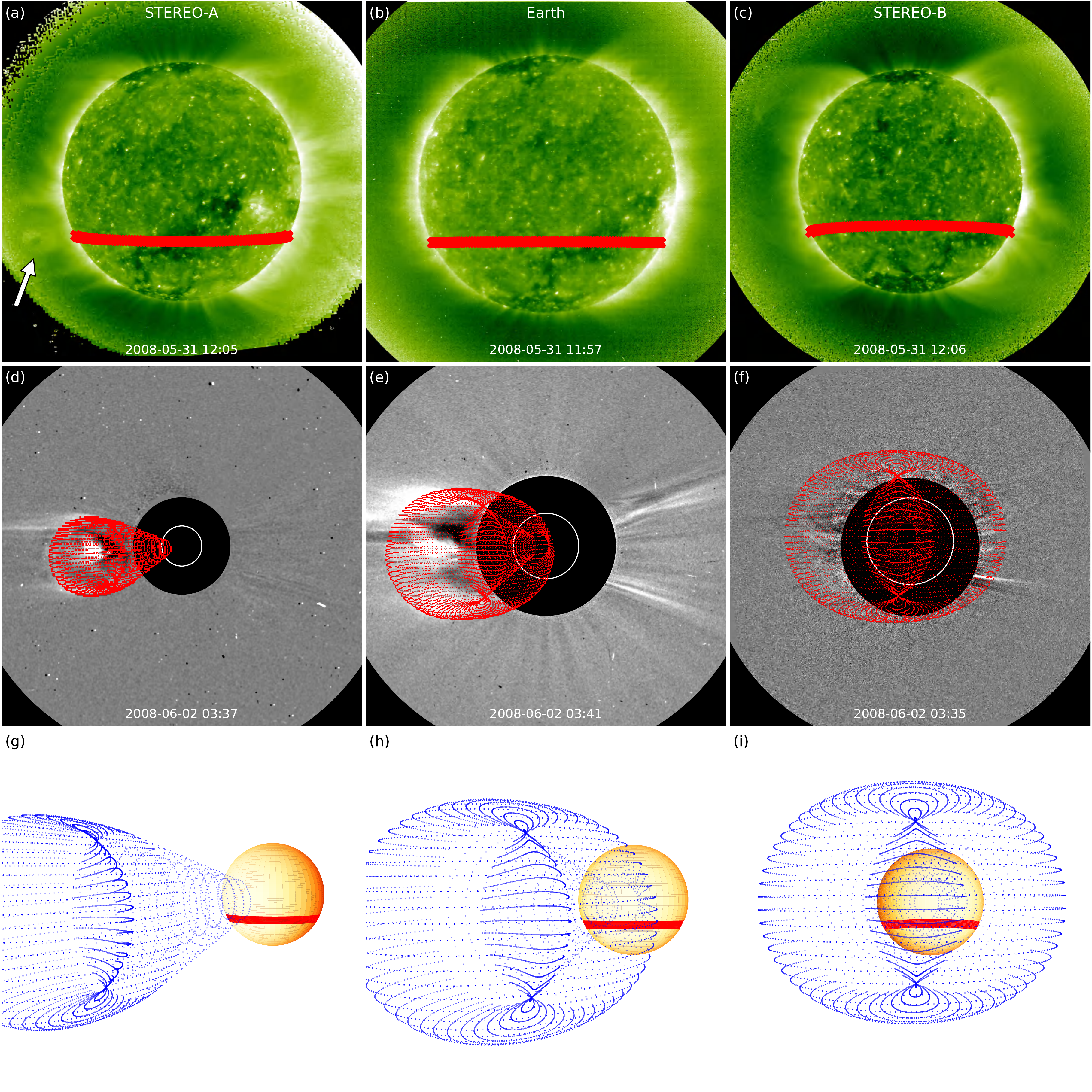}
\end{center}
\caption{Geometric techniques applied to the 2008 June 1 CME (Event 1). (a--c) Latitude (thick red line) of the CME core (indicated with an arrow) projected onto the solar disc as seen by STEREO-A, SOHO, and STEREO-B. (d--f) GCS reconstruction applied to the white-light structure seen in STEREO-A, SOHO, and STEREO-B imagery. (g--i) The GCS-reconstructed CME displayed together with the `source latitude' in a 3D representation and shown from the three viewpoints of STEREO-A, Earth, and STEREO-B.}
\label{fig:20080601_geometric}
\end{figure}

Results from the application of geometric techniques to Event~1 are shown in Figure~\ref{fig:20080601_geometric}. In the top row, the approximate latitude of the footpoints of the flux rope structure seen in off-limb imagery from STEREO-A (marked with an arrow in Figure~\ref{fig:20080601_geometric}(a), see also Supplementary Video~1) is projected onto the solar disc from all three available viewpoints. The flux rope lifted off from ${\sim}$S27$^\circ$, roughly consistently with the location of the faint dimmings shown in Figure~\ref{fig:20080601_imaging}, which extend between ${\sim}$S20$^\circ$ and ${\sim}$S50$^\circ$. We also perform a GCS reconstruction of the large-scale CME in the corona, shown in the middle row of Figure~\ref{fig:20080601_geometric}. We remark that, given that the eruption is a halo from STEREO-B and rather close to the central meridian as seen from Earth, the first available time for a meaningful fitting is ${\sim}$3.5~hours after the first appearance of the CME in the COR2-A field of view. According to GCS results, the CME apex is located at S04$^\circ$E29$^\circ$ as seen from Earth, which converts into a longitude of E04$^\circ$ as seen from STEREO-B. Whilst the value for latitude is significantly different, the longitude is quite consistent with that of the faint dimmings indicated in the last column of Figure~\ref{fig:20080601_imaging}, which cover a longitudinal span of E25$^\circ$--W25$^\circ$ in the STEREO-B reference frame (note that the images in the middle row of Figure~\ref{fig:20080601_geometric} and those in the last column of Figure~\ref{fig:20080601_imaging} are taken at the same time). The `source latitude' and the reconstructed GCS wireframe are shown together in a 3D representation in the bottom row of Figure~\ref{fig:20080601_geometric}. It is clear that the CME deflected significantly towards the solar equatorial plane during its early evolution, which is also visible from Supplementary Video~1 and is consistent with the tendency of CMEs to align themselves with the heliospheric current sheet during solar minimum \citep[e.g.,][]{yurchyshyn2009,isavnin2014}. Hence, we conclude that searching for the source of the eruption based on coronagraph images and GCS reconstructions alone would have likely resulted in a somewhat misleading region.

\subsection{Event 2: 2011 March 3} \label{subsec:2011mar}

The second CME that we analyse in this work (Event 2) erupted on 2011~March~3 and was a stealth event as seen from Earth. This case study was previously analysed by \citet{pevtsov2012}, \citet{nitta2017}, \citet{okane2019}, and \citet{okane2021a}, who all placed the CME source region in the vicinity of active region AR~11165, located close to the central meridian from Earth's perspective. Hence, we investigate this event as a possible case of a more localised, active region stealth CME. At the time of this eruption, STEREO-A was located $87^{\circ}$ west of Earth and STEREO-B was positioned $95^{\circ}$ west of Earth, meaning that the two spacecraft had a nearly-quadrature view of the event from opposite sides. Such observations are shown in Supplementary Video~3, which presents simultaneous EUVI and COR1 data from the twin STEREOs. In the video, a large bubble-shaped set of loops can be initially seen to lie off the limb above AR~11165, before slowly inflating (starting around 18:00~UT on March~2) and erupting as a flux rope CME that reached the COR2 field of view around 04:00~UT on March~3. From the STEREO-A perspective, high-altitude post-eruption arcades could be observed after the CME lifted off as AR~11165 rotated into view, forming just above the longer-lived active region loops. It is clear from these observations alone that this CME indeed originated from AR~11165, albeit from higher altitudes than usual and thus away from stronger active region fields. This may be a contributing factor to the stealthiness of the event in on-disc imagery, as suggested by \citet{okane2019}.

\begin{figure}[!ht]
\begin{center}
\includegraphics[width=0.99\linewidth]{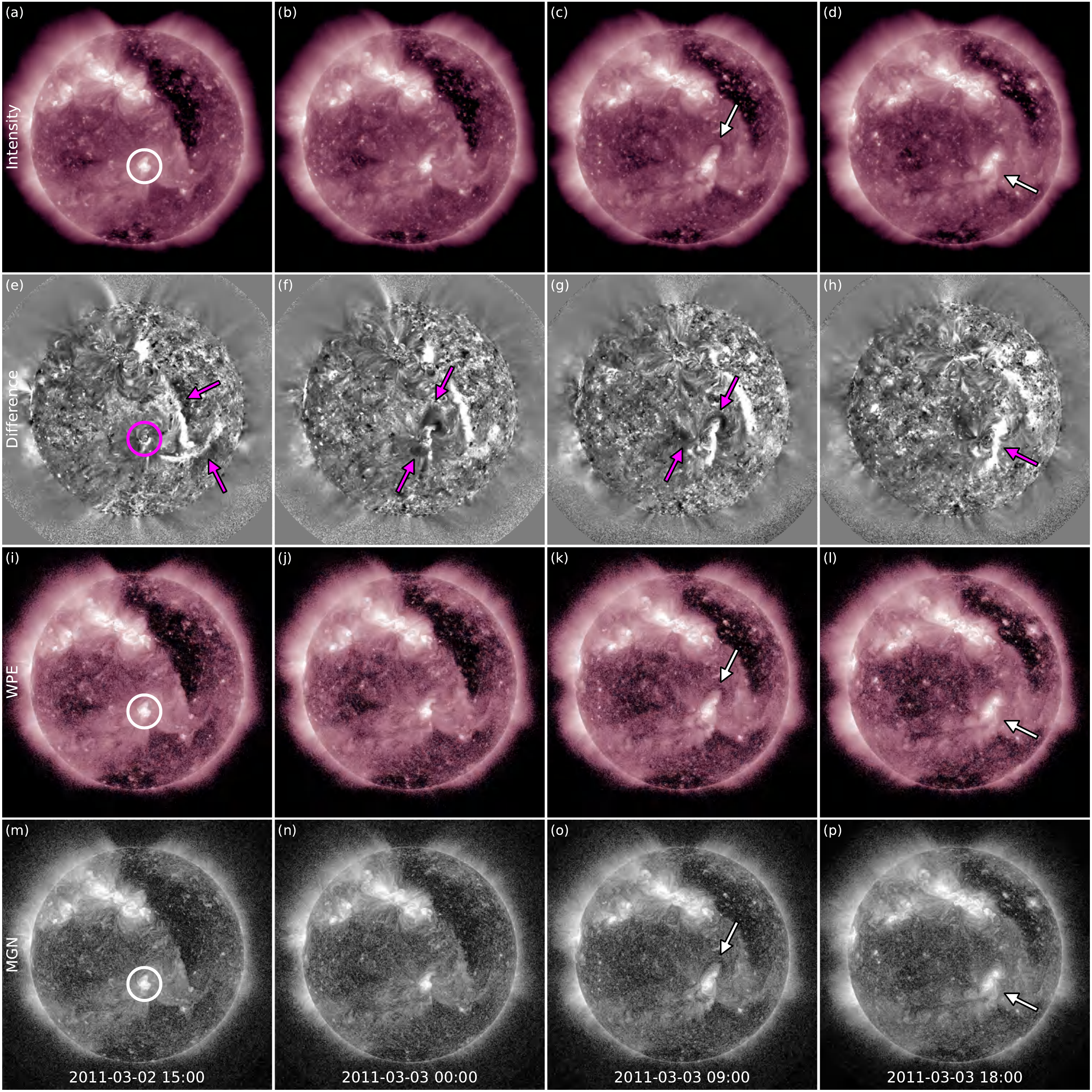}
\end{center}
\caption{Imaging techniques applied to the 2011 March 3 CME (Event 2). SDO/AIA 211~{\AA} images are shown at four different times and processed with four different methods. (a--d) Intensity images. (e--h) Difference images with fixed $\Delta t = 12$~hours. (i--l) WPE-processed images. (m--p) MGN-processed images. AR~11165 is circled in the first column. Dimming and brightening regions are indicated with arrows (see text for details).}
\label{fig:20110303_imaging}
\end{figure}

Images from the Atmospheric Imaging Assembly \citep[AIA;][]{lemen2012} instrument onboard SDO corresponding to the eruption period are shown in Figure~\ref{fig:20110303_imaging} and Supplementary Video~4. Whilst the intensity, MGN, and WPE images do not show strong eruptive signatures, the difference images in the second row display clearer signs of activity on the solar disc. First of all, an extended pair of dimmings and flare loops at the southern periphery of a large coronal hole in the western hemisphere can be observed in Figure~\ref{fig:20110303_imaging}(e) (indicated with arrows). These are found to be associated with a filament eruption that occurred on March~1, which possibly destabilised the nearby active region fields and thus triggered the later stealth CME. The successive panels are characterised by a series of dimmings (starting around 22:00~UT on March~2 and marked in panels (f) and (g)) and a strong brightening (starting around 01:00~UT on March~3 and marked in panel (h)) around AR~11165 (indicated with a circle in the first column of Figure~\ref{fig:20110303_imaging}). It can be seen that initially, a pair of dimmings develops north--south of the active region and later a third, more diffuse dimming appears to the east of the fading southern one. Corresponding images obtained with other techniques, on the other hand, show more elusive signatures. A faint pair of dimmings north and south of AR~11165 can be seen in Supplementary Video~4 (the northern dimming is marked in panels (c), (k), and (o) of Figure~\ref{fig:20110303_imaging}), together with the appearance of a set of loops that seem to correspond to the post-eruption arcade seen by STEREO-A (marked in panels (d), (l), and (p) of Figure~\ref{fig:20110303_imaging}). We emphasise that these signatures are rather weak, and thus without prior knowledge of an eruption having occurred they would have been easily overlooked. Furthermore, we note that although the arcade is well observed in all non-differenced data (i.e., intensity, WPE, and MGN), its fine structure is more easily revealed in images processed with advanced techniques.

\begin{figure}[t!]
\begin{center}
\includegraphics[width=0.75\linewidth]{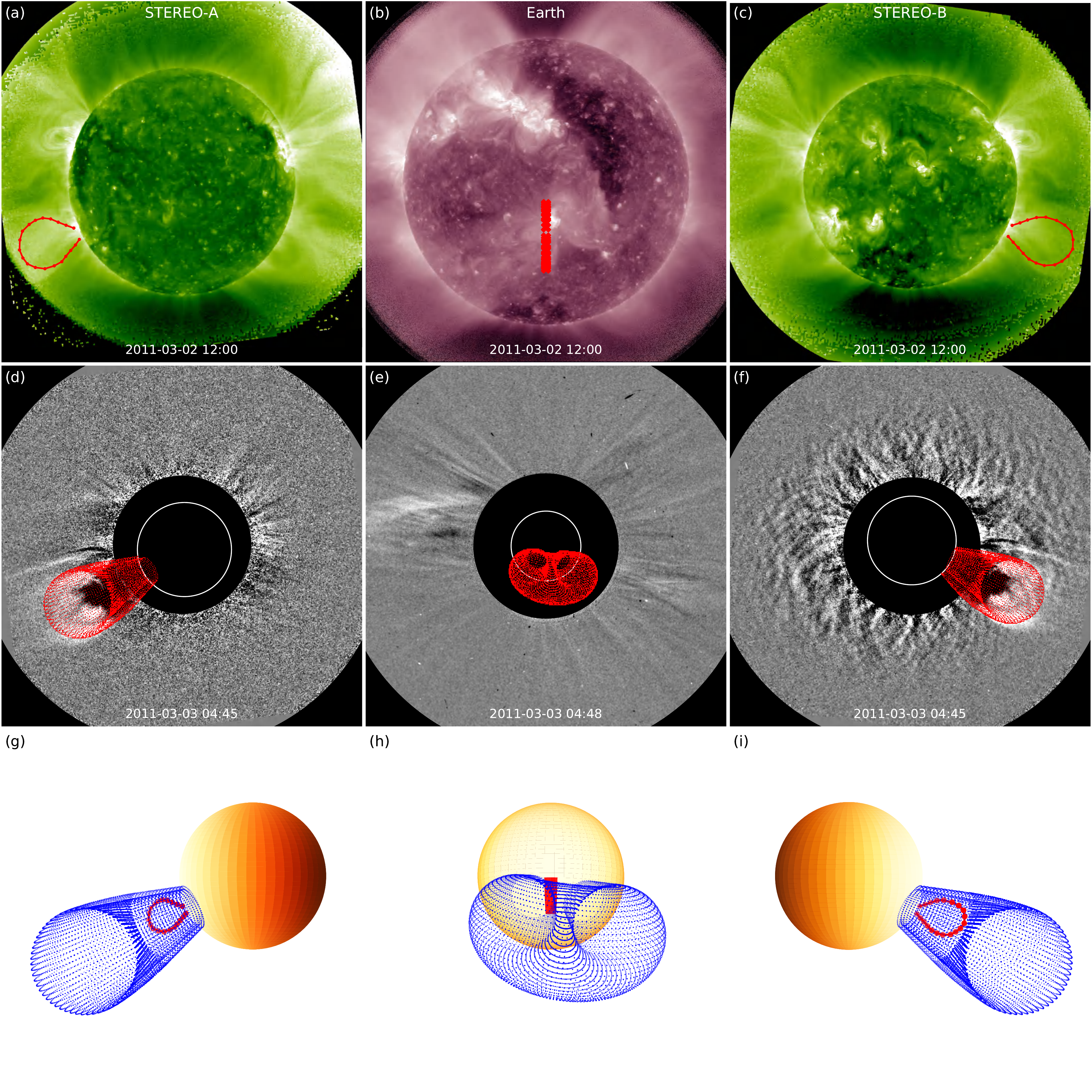}
\end{center}
\caption{Geometric techniques applied to the 2011 March 3 CME (Event 2). (a--c) Application of the TP technique to the pre-eruptive structure. The technique is applied to STEREO-A and STEREO-B images and the latitudinal extent of the triangulated loop is projected onto the SDO/AIA field of view (and displayed over the central meridian). (d--f) GCS reconstruction applied to the white-light structure seen in STEREO-A, SOHO, and STEREO-B imagery. (g--i) Results of the TP and GCS reconstruction techniques displayed together in a 3D representation and shown from three viewpoints: STEREO-A, Earth, and STEREO-B.}
\label{fig:20110303_geometric}
\end{figure}

Results from the application of geometric techniques to Event~2 are shown in Figure~\ref{fig:20110303_geometric}. Since the pre-eruptive structure was well observed by both STEREO spacecraft, we analyse it using the TP technique. The triangulated balloon-shaped feature is overlaid onto EUVI images in panels (a) and (c). However, despite the favourable viewing perspective of nearly-quadrature with Earth from both observers, it is not possible to obtain a meaningful triangulation of the loop onto the solar disc imaged by SDO. This is because the uncertainty associated with the TP technique depends on the angle between the observing spacecraft, and specifically is proportional to the inverse of the sine of the separation angle \citep{inhester2006}. In the case of this event, the separation of ${\sim}180^{\circ}$ between the two STEREOs results in a large uncertainty in the east--west direction.  Hence, the only information that can be retrieved from the TP technique applied to Event~2 is the latitudinal position of each point part of the triangulated loop, resulting in a vertical bar in Figure~\ref{fig:20110303_geometric}(b) that is arbitrarily placed at central meridian relative to Earth since its exact longitude cannot be retrieved. The extent of the loop in the north--south direction is consistent with the location of AR~11165, as expected. Results of the GCS reconstruction are shown in the middle panels of Figure~\ref{fig:20110303_geometric}. Since the CME was well observed as a limb event from both STEREO spacecraft, we perform our reconstruction while the transient is still visible in SECCHI/COR1 and before it reaches the LASCO/C2 field of view, in order to capture the eruption as close as possible to its initial state. This results in a propagation direction of S27$^\circ$W07$^\circ$ as seen from Earth, rather close to location of AR~11165 at the reconstruction time (S17$^\circ$W12$^\circ$), but still suggesting a slight southward deflection. Results from the two geometric techniques are shown together in a 3D representation in the bottom panels of Figure~\ref{fig:20110303_geometric}. Despite the issues with the application of the TP technique for this particular event discussed above, the two structures obtained match fairly well. Even if an exact longitude of the source region cannot be retrieved from TP, it may be argued that the viewing geometry of the two STEREO spacecraft implies that the triangulated structure was located close to central meridian. An alternative would be to perform a GCS reconstruction based on the EUV images, as made e.g.\ by \citet{okane2021b}.

\subsection{Event 3: 2012 February 4} \label{subsec:2012feb}

The third CME that we analyse in this work (Event 3) erupted on 2012~February~4 and was a stealth event as seen from Earth. This CME was first reported by \citet{dhuys2014} and was further analysed by \citet{alzate2017}. This event was also well observed off limb by both STEREO spacecraft, with STEREO-A being located $108^{\circ}$ west of Earth and STEREO-B being placed $115^{\circ}$ east of Earth. Observations from both EUVI and COR1 telescopes are provided in Supplementary Video~5 and show the CME of interest being ejected off the northeastern limb from STEREO-A's viewpoint and off the northwestern limb from STEREO-B's perspective, indicating that the eruption originated from the Earth-facing Sun. The pre-eruptive structure could be observed for several hours above the limb from both spacecraft before its slow lift-off (starting around 04:00~UT on February~4), indicating that the CME flux rope erupted from unusually high altitudes. The CME reached the COR2 field of view around 10:00~UT on February~4.

\begin{figure}[t!]
\begin{center}
\includegraphics[width=0.99\linewidth]{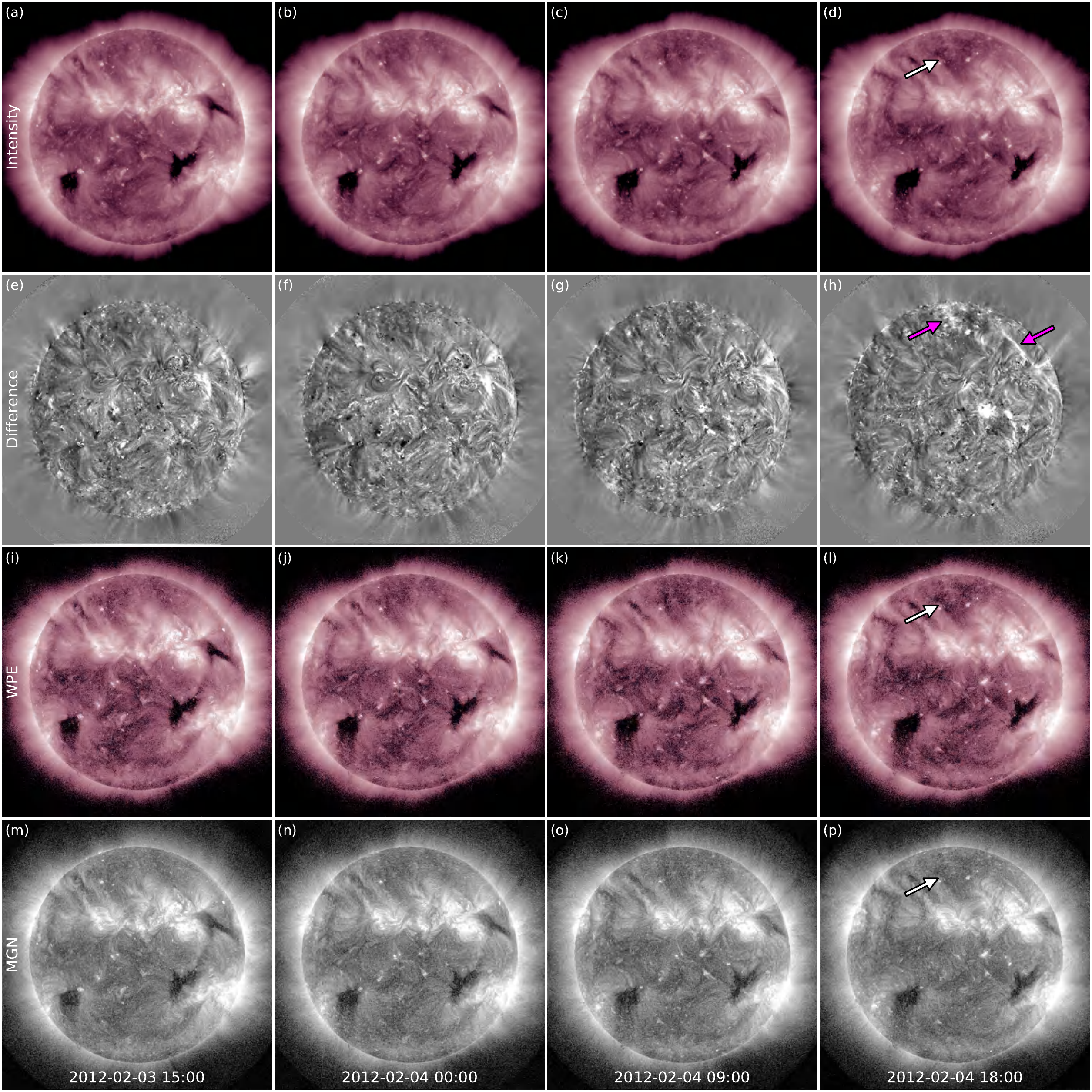}
\end{center}
\caption{Imaging techniques applied to the 2012 February 4 CME (Event 3). SDO/AIA 211~{\AA} images are shown at four different times and processed with four different methods. (a--d) Intensity images. (e--h) Difference images with fixed $\Delta t = 12$~hours. (i--l) WPE-processed images. (m--p) MGN-processed images. Dimming and brightening regions are indicated with arrows (see text for details).}
\label{fig:20120204_imaging}
\end{figure}

SDO/AIA images corresponding to the eruption period are shown in Figure~\ref{fig:20120204_imaging} and Supplementary Video~6. As is often the case for stealth CMEs, these data do not show ``explosive''  eruption signatures that are more typical of active-region CMEs. Nevertheless, it is possible to distinguish rather clear indications that an eruption occurred, as marked with arrows in the difference image in Figure~\ref{fig:20120204_imaging}(h). Specifically, we observe an elongated pair of brightenings attributable to flare ribbons, the first being rather prominent and close to the northwestern limb and the second being more diffuse and culminating in the vicinity of the solar north pole, whit faint loops reminiscent of a post-eruption arcade in between. These features start to develop around 12:00~UT on February~4. In non-differenced images, the westernmost ribbon appears significantly less prominent and the more diffuse one is not visible at all. The structure that we recognised as a post-eruption arcade based on difference images is somewhat visible, but less clearly attributable to a post-eruption arcade. Although all these coronal features can be observed in intensity images as well, their structure appears sharper in WPE- and MGN-processed data. Furthermore, all images reveal an especially faint dimming region close to the solar north pole (with onset around 06:00~UT on February~4 and indicated with arrows in Figure~\ref{fig:20120204_imaging}(d), (l), and (p)). This darkening feature appears equally visible in intensity, MGN, and WPE images. Supplementary Video~6 demonstrates that while it is not possible to establish with certainty whether a large-scale eruption occurred from intensity images alone, the development of the various brightening features in difference images unambiguously links the CME seen off limb from the twin STEREOs to a rather defined source region on the Earth-facing disc. 

\begin{figure}[t!]
\begin{center}
\includegraphics[width=0.75\linewidth]{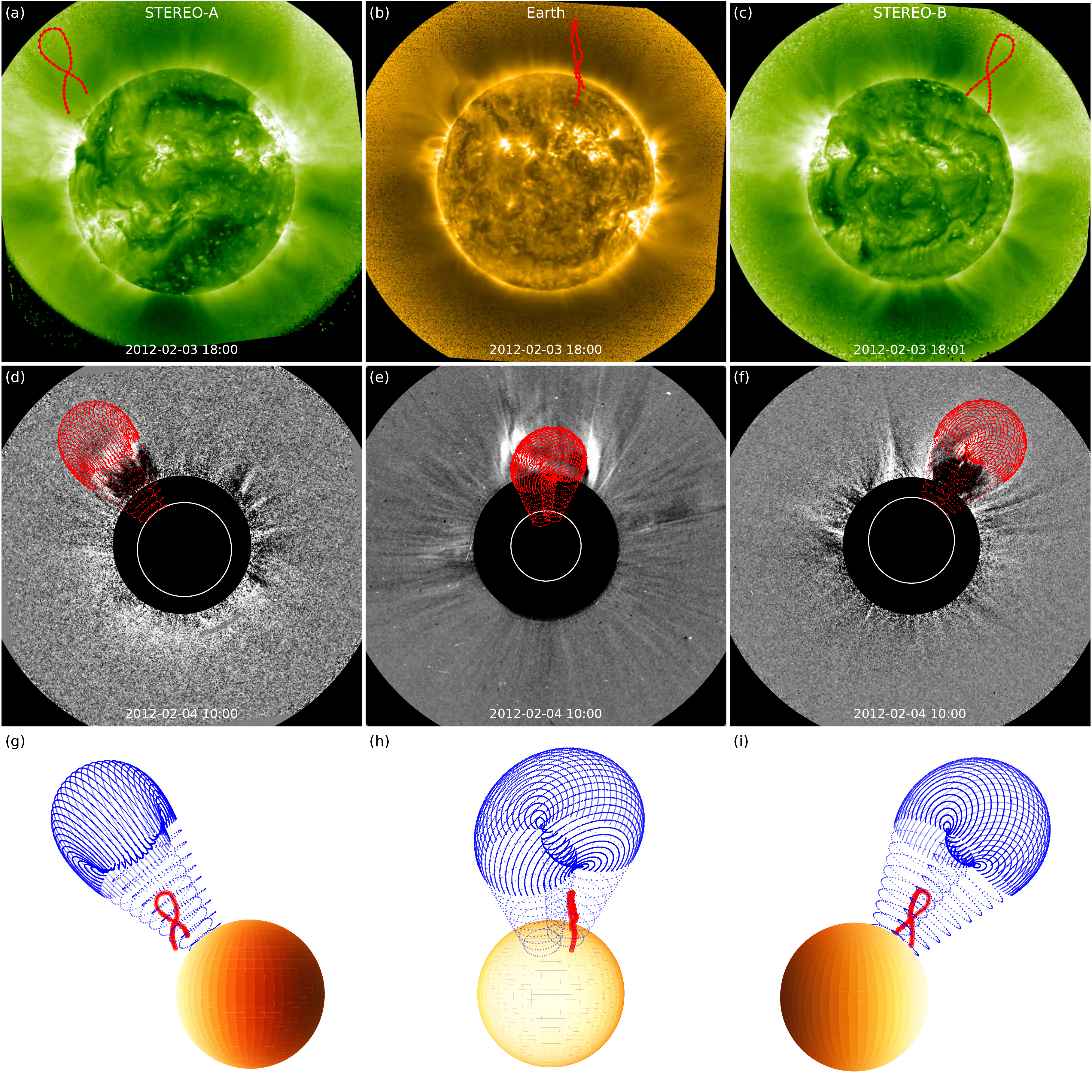}
\end{center}
\caption{Geometric techniques applied to the 2012 February 4 CME (Event 3). (a--c) Application of the TP technique to the pre-eruptive loop structure. The technique is applied to STEREO-A and STEREO-B images and then projected into the PROBA2 field of view. (d--f) GCS reconstruction applied to the white-light structure seen in STEREO-A, SOHO, and STEREO-B imagery. (g--i) Results of the TP and GCS reconstruction techniques displayed together in a 3D representation and shown from the viewpoints of STEREO-A, Earth, and STEREO-B.}
\label{fig:20120204_geometric}
\end{figure}

Figure~\ref{fig:20120204_geometric} shows the results of the geometric techniques applied to Event~3. Being well-observed off limb from two different perspectives, the 2012~February~4 CME is well suited to be analysed using the TP technique. Furthermore, the ${\sim}135^{\circ}$ separation between the two STEREO spacecraft does not lead to the uncertainty issues that were encountered for Event~2. In the top row of Figure~\ref{fig:20120204_geometric}, a pre-eruptive loop (on 2012~February~3 at 18:00~UT) is traced with the TP technique in 195~{\AA} images from both STEREO spacecraft and then projected onto Earth's view, resulting in a structure rooted around N55$^{\circ}$W30$^{\circ}$. Note that the triangulated loop has been projected into a 174~{\AA} image from the Sun Watcher using Active Pixel System Detector and Image Processing \citep[SWAP;][]{halain2013,seaton2013} telescope onboard the Project for On Board Autonomy 2 \citep[PROBA2;][]{santandrea2013} satellite, which has a larger field of view than SDO/AIA. The highest point in the reconstructed loop lies at $1.81\,R_{\odot}$ from the solar centre, confirming that a high-altitude flux rope was involved in the eruption. The middle row of Figure~\ref{fig:20120204_geometric} shows the GCS technique applied to simultaneous images of the white-light CME from the SECCHI/COR2 coronagraphs onboard STEREO and the LASCO/C2 coronagraph onboard SOHO. The CME apex has the direction N50$^{\circ}$W03$^{\circ}$, consistent with a high-latitude eruption. Results from the two reconstructions are combined in the bottom row of Figure~\ref{fig:20120204_geometric}. Whilst the latitudes retrieved from the two methods are in agreement, larger differences are found in the longitudes, which is not surprising considering, for example,\ that the GCS precision is typically around $\pm4^{\circ}$ for latitude and $\pm17^{\circ}$ for longitude \citep{thernisien2009}. Nevertheless, both techniques yield a source region that is (at least to some extent) consistent with the bright features indicated in Figure~\ref{fig:20120204_imaging}(h).

\subsection{Event 4: 2016 October 9} \label{subsec:2016oct}

The fourth CME that we analyse in this work (Event~4) erupted on 2016 October 8 and was a stealth event as seen from Earth. This CME was first reported by \citet{nitta2017} and was further analysed by \citet{he2018}. Having taken place in 2016, only the viewpoint from STEREO-A located $148^{\circ}$ east of Earth is available since contact with STEREO-B was lost in October 2014. Observations from the EUVI and COR1 telescopes onboard STEREO-A are presented in Supplementary Video~7, where the eruption can be seen to originate close to the solar equator off the western limb, corresponding to an ejection off the Earth-facing disc. In addition, the CME cavity extended significantly beyond the central core early in the eruption, suggestive of a much larger-scale event. Loops corresponding to the outer CME rim can be observed to lift off ${\sim}25^{\circ}$ south of the equator around 16:00~UT on October~8, whilst structures to the north of the central core cannot be discerned with clarity in the EUVI field of view, most likely because of the presence of a bright helmet streamer. The resulting CME reached the COR2 field of view around 22:00~UT on October~8.

\begin{figure}[!ht]
\begin{center}
\includegraphics[width=0.99\linewidth]{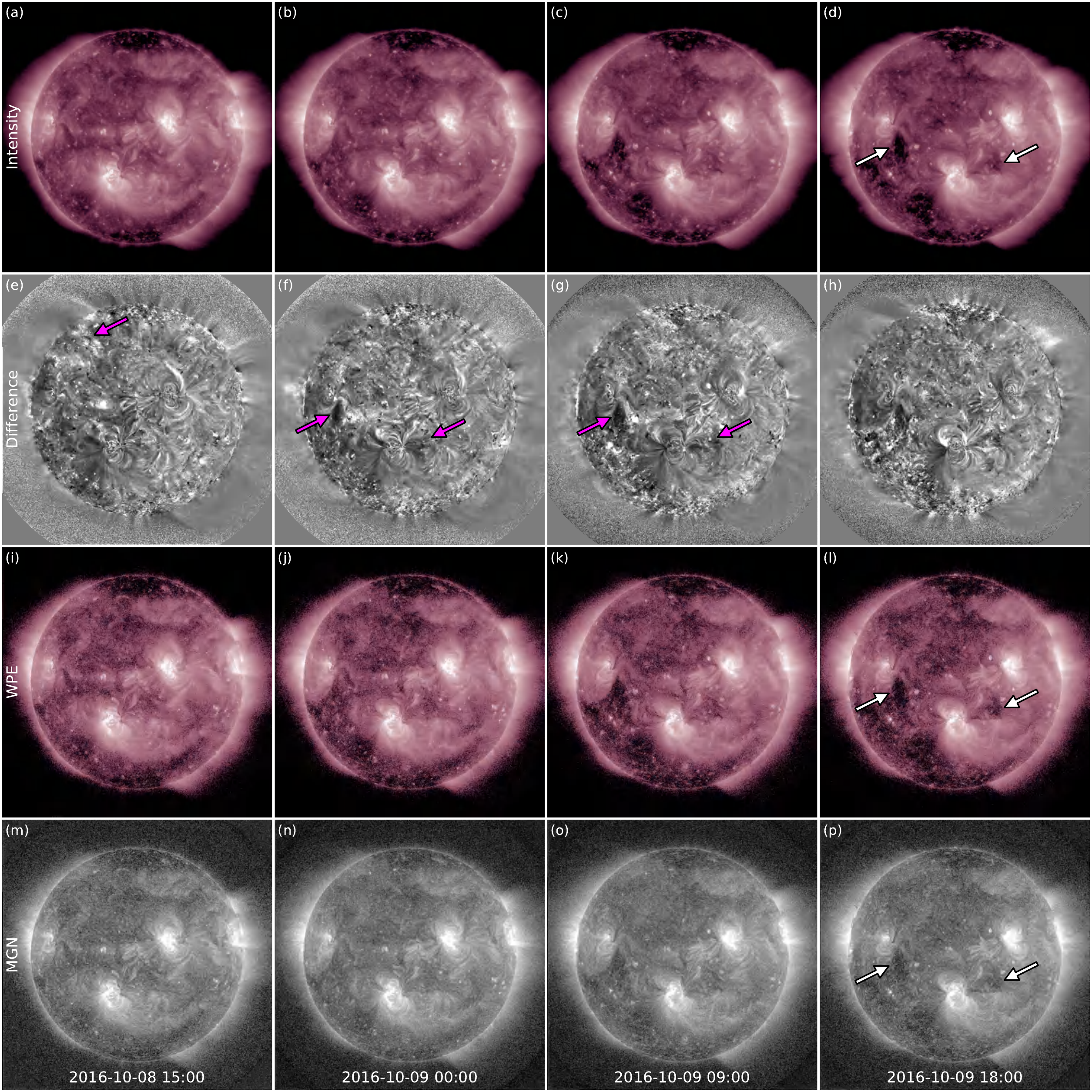}
\end{center}
\caption{Imaging techniques applied to the 2016 October 8 CME (Event 4). SDO/AIA 211~{\AA} images are shown at four different times and processed with four different methods. (a--d) Intensity images. (e--h) Difference images with fixed $\Delta t = 12$~hours. (i--l) WPE-processed images. (m--p) MGN-processed images. Dimming and brightening regions are indicated with arrows (see text for details).}
\label{fig:20161008_imaging}
\end{figure}

Images from SDO/AIA corresponding to the eruption period are shown in Figure~\ref{fig:20161008_imaging} and Supplementary Video~8. It is possible to discern signatures of two different eruptions in the presented data. The first can be noted clearly in the figure in difference images (indicated with an arrow in panel (e)) and in all panels in Supplementary Video~8, and corresponds to a small filament eruption from the northeastern quadrant. Despite the rather evident low-coronal signatures, this eruption does not correspond to the CME seen off limb in STEREO-A imagery (see Supplementary Video~7) because of its timing ($\sim$14:00~UT on October~8, several hours too early compared to its first appearance in COR1), its source region (${\sim}40^{\circ}$ north of the equator), and its localised nature (in contrast to the large-scale CME observed by STEREO-A). Nevertheless, it is possible that this minor eruption destabilised the overlying field(s) and thus facilitated the onset of the following, larger CME. Over the following hours, starting around 18:00~UT on October~8, a pair of coronal dimmings developed, marked by arrows in Figure~\ref{fig:20161008_imaging} (d) and evident in Supplementary Video~8. In the difference images in panels (f) and (g), it is evident that the eastern dimming appears deeper than the western one. In data processed with WPE or MGN in panels (i) to (l) and (m) to (p), respectively, the progressive darkening of these two areas over the presented interval shows that the dimmings developed over a remarkably long time span (i.e., even longer than the $\Delta t = 12$~hours used here for difference images). We do not note differences in the appearance of the dimmings in the intensity, WPE, and MGN images. Their extent in latitude (from ${\sim}$N02$^{\circ}$ to ${\sim}$S22$^{\circ}$) is in agreement with the location of the off-limb signatures observed from the STEREO-A viewpoint and shown in Supplementary Video~8.

\begin{figure}[t!]
\begin{center}
\includegraphics[width=0.5\linewidth]{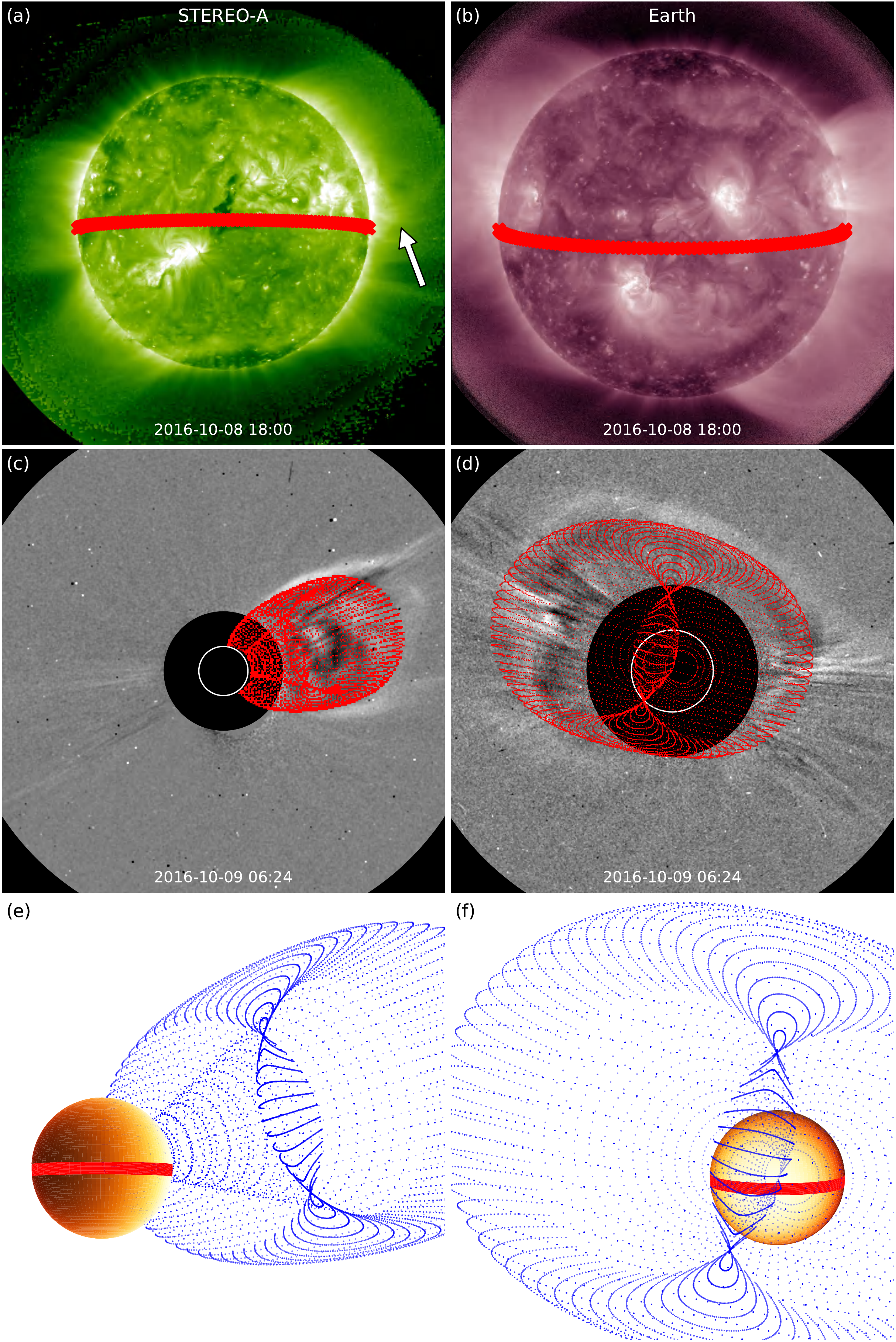}
\end{center}
\caption{Geometric techniques applied to the 2016 October~8 CME (Event~4). (a--b) Latitude (thick red line) of the CME core (indicated with an arrow) projected onto the solar disc as seen by STEREO-A and SDO. (c--d) GCS reconstruction applied to the white-light structure seen in STEREO-A and SOHO. (e--f) The GCS-reconstructed CME displayed together with the `source latitude' from (a) and (b) in a 3D representation and shown from the two viewpoints of STEREO-A and Earth.}
\label{fig:20161008_geometric}
\end{figure}

Results from applying geometric techniques to Event~4 are shown in Figure~\ref{fig:20161008_geometric}. In the top row, the red line indicates the approximate latitude of the CME core as derived from STEREO-A observations (marked with an arrow in Figure~\ref{fig:20161008_geometric}(a), see also Supplementary Video~7) projected onto the solar disc from both available viewpoints. The latitude of ${\sim}$S02$^{\circ}$ is rather consistent with the dimming locations, but less compatible with the GCS reconstruction shown in the middle row of Figure~\ref{fig:20161008_geometric}, which gives a CME apex propagation direction of N12$^{\circ}$E05$^{\circ}$, indicating a significant deflection and/or non-radial propagation of the structure towards the north after eruption (this aspect can also be noted in Supplementary Video~8). The dimmings covered a longitudinal span of E44$^{\circ}$--W22$^{\circ}$ at the time of the GCS reconstruction, highlighting the large-scale nature of the event, and were centred around E12$^{\circ}$, i.e.,\ well within the GCS uncertainties mentioned in Section~\ref{subsec:2012feb}. As was the case for Event~2 and Event~3, this eruption can be convincingly linked to low-coronal signatures (albeit weak), and as was the case for Event~1, relying uniquely on coronagraph imagery and GCS reconstructions would have resulted in a somewhat misleading estimated source region. The fact that this CME was a full halo as seen by SOHO and that only two viewpoints were available certainly contributed to the late reconstruction time, when the apex was already at ${\sim}10$\,$R_{\odot}$. Hence, the geometric parameters were only determined after the CME had already experienced significant alterations to its trajectory.


\section{Discussion and Conclusions} \label{sec:conclusions}

In this work, we have presented and analysed four stealth CMEs that presented diverse characteristics: a classic streamer blowout (Event 1), a CME originating from an active region (Event 2), a flux rope lying at unusually high altitudes prior to eruption (Event 3), and a significantly large-scale event (Event 4). These case studies were also characterised by different viewing geometries between the ``stealthy perspective'' (on the solar disc) and the off-limb observer(s): Event~1 has STEREO-A ${\sim}50^{\circ}$ away from STEREO-B, Event~2 had the STEREOs nearly in quadrature with Earth, Event~3 had the STEREOs ${\sim}110^{\circ}$ away from Earth, and Event~4 had STEREO-A separated by ${\sim}150^{\circ}$ from Earth. We have investigated these CMEs using remote-sensing imaging and geometric techniques in order to determine their corresponding source region on the Sun. Our analysis was based on EUV images of the solar disc in the 195~{\AA} (STEREO) and 211~{\AA} (SDO) channels, especially suited to detect dimming and brightening regions associated with low-coronal signatures of CMEs \citep[e.g.,][]{nitta2017}, together with white-light images of the solar corona. Since for all events an approximate region of origin was known due to off-limb views from additional viewpoints, the motivation for our analysis was to test and demonstrate the effectiveness of the different techniques over a range of events with different properties and observation geometries. Our main findings are summarised in Table~\ref{tab:results}.

{
\renewcommand{\arraystretch}{1.2}
\begin{table}[ht]
\center
\caption{Results from applying different imaging and geometric techniques to the events under study. For the imaging techniques, the observed eruption signatures are indicated, with A = arcade, B = brightening, and D = dimming. The upper cases denote strong signatures, whilst the lower cases denote only weak signatures. For the geometric techniques, it is indicated whether the TP method could be applied (N/A for events that only had one off-limb view available) and which features could be triangulated, as well as the radial distance from the Sun of the CME apex at which the earliest meaningful GCS reconstruction could be performed. The `latitude projection' scheme could in principle be applied to all events under study, since it only requires one off-limb viewpoint of the eruption, hence it is not shown here.} \label{tab:results}
\vspace*{.1in}
\begin{tabular}{l @{\hskip .2in} l @{\hskip .35in} c @{\hskip .2in} c @{\hskip .2in} c @{\hskip .2in} c @{\hskip .4in} c @{\hskip .2in} c @{\hskip .2in} c @{\hskip .2in} c}
\toprule
& & \multicolumn{4}{c}{\textbf{Imaging}} & \multicolumn{2}{c}{\textbf{Geometric}}\\
\cmidrule(r{15pt}){3-6} \cmidrule(r){7-8}
& & \textbf{Intens} & \textbf{Diff} & \textbf{MGN} & \textbf{WPE} & \textbf{TP} & \textbf{GCS}\\
\midrule
\textbf{E1} & Streamer blowout & d & d & d & d & N/A & ${\sim}7\,R_{\odot}$\\
\textbf{E2} & AR eruption & A,d & a,B,D & A,d & A,d & Lat only & ${\sim}3\,R_{\odot}$ \\
\textbf{E3} & High flux rope & b,d & A,B,d & b,d & b,d & Full loop & ${\sim}4\,R_{\odot}$ \\
\textbf{E4} & Large-scale CME & D & D & D & D & N/A & ${\sim}10\,R_{\odot}$\\
\bottomrule
\end{tabular}
\end{table}
}

Analysis of solar disc imagery with various image processing techniques revealed the presence of signatures for all events, as shown in Table~\ref{tab:results}, albeit with different confidence levels. The most convincing evidence was found for Event~2 and Event~3, whilst Event~4 was associated with weak but reasonable signatures and Event~1 was characterised by the largest uncertainties. At least according to the small sample investigated here, it seems that the main factor contributing to the level of ``stealthiness'' is the spatial extent of the eruption rather than the altitude from which a CME lifts off (in particular, compare the size of the on-disc signatures for Event~1 and Event~4 with those for Event~2 and Event~3, which are significantly more localised). Another aspect to note is that the eruption associated with the weakest signatures (Event~1) was also the only event that did not feature a nearby active region or area of strong magnetic fields. Regarding the image processing techniques used, it is evident that difference images with large temporal separations revealed the clearest eruptive signatures, as was also reported by \citet{nitta2017}. Moreover, we note that even if dimmings are often evident enough in intensity data, brightenings tend to appear overwhelmingly clearer in difference images (see Event~2 and Event~3). In this work, we have used a fixed $\Delta t = 12$~hours for all events, but even longer separations may be explored in the case of eruptions that develop extremely slowly (as for Event~1 and Event~4), although artefacts at the solar limb would also become more prominent and problematic. In this regard, it should be noted that difference images are particularly prone to spurious effects due to spatio-temporal interference; i.e., dimming and brightening features may correspond to ``true'' dimmings and brightenings as well as moving structures over long time scales. Thus, difference images may be complemented with non-differenced data, which should be used to properly interpret the identified large-scale changes and connect them to well-defined activity on the Sun. Furthermore, we did not find substantial differences in the features revealed in  ``normal'' intensity images and those produced by more advanced processing techniques for the cases that were only associated with (more or less defined) dimmings, i.e. Event~1 and Event~4. Since the main purpose of these methods is to sharpen coronal features and accentuate small-scale variations, it is not surprising that they are as powerful as non-processed images when such structures and alterations are missing in the first place, i.e.\ in the case of the most problematic events. This overall conclusion is in agreement with \citet{okane2021b}, who studied a stealth CME off limb (from a secondary viewpoint) using MGN, but found that this technique did not also reveal the corresponding signatures on disc, which were however completely elusive to other data sets as well. On the other hand, the events that were associated with more prominent structural changes in the corona, i.e.\ Event~2 and Event~3, showed significantly enhanced features in WPE and MGN imagery in comparison to intensity data. This allows for deeper analysis of the onset and signatures of these eruptions, which may in turn advance understanding of at least a subset of elusive events. Hence, this work demonstrates that advanced image processing techniques are also applicable to a portion of large-scale stealth CMEs observed against the solar disc, in addition to their usefulness for investigating small, short-lived activity and off-limb events that has been shown in previous studies \citep[e.g.,][]{alzate2017,okane2019,liewer2021}. In conclusion, our recommendation for identifying and analysing the origins of stealth CMEs on the solar disc is a multi-step approach: (1) use difference images to easily single out large-scale changes, (2) use intensity data to properly interpret difference images and rule out artefacts, and (3) if coronal features such as brightenings, loops, and ribbons can be identified, use advanced image processing techniques such as WPE and MGN to analyse their fine structure in deeper detail.

Analysis of the events in this study with the aid of various geometric techniques has revealed that triangulation and reconstruction methods can help trace an eruption back to its source as long as they are used when the CME is as close as possible to the Sun. Given that most CME deflections and other non-radial propagation effects take place below a few solar radii \citep[e.g.,][]{kay2015a,kay2015b,liewer2015}, it is crucial to determine the geometric parameters before the most dramatic evolution has occurred. In this sense, the CME propagation direction with respect to the observers plays a central role, as can be seen from the reconstructed CME apex heights shown in Table~\ref{tab:results}: amongst the cases investigated here, the best scenario was achieved for Event~2 and Event~3, in which the CMEs were propagating in directions well away from at least two viewpoints, enabling a meaningful GCS reconstruction to be made early on. The least favourable configuration happened for Event~4, where the CME was a full halo from one viewpoint and only a second observer was available, thus the apex was already at ${\sim}10$\,$R_{\odot}$ at the time of the performed GCS fitting. In this case, the resulting CME reconstruction strongly hints at an eruption that originated from the northern hemisphere. Hence, without complementary observations from STEREO-A, the source of the white-light structure might have been erroneously attributed to the previous small filament eruption preceding the ``main'' event. Hence, our recommendation for tracing a CME observed in white light back to its elusive source is to mind the viewing configuration of the event and pay caution the farther the reconstructed CME is from the Sun. The TP technique, on the other hand, is quite efficient in triangulating a pre-eruptive structure and/or tracing an eruption back to a source, but is more strongly dependent on the viewing geometry (see Table~\ref{tab:results}). Excluding cases in which a spacecraft separation of ${\sim}180^{\circ}$ does not allow for a unique solution (see e.g.\ Event~2), the method requires two well-separated spacecraft to observe a particular feature simultaneously, so it is not applicable to cases in which an eruption is behind the limb relative to at least one observer. Of course, being able to analyse a stealth CME with triangulation and reconstruction methods implies the availability of additional viewpoints to start with. Apart from helping to discern whether a CME is stealthy or simply far-sided, remote-sensing measurements away from the Sun--observer line may help identify the source of an elusive event in a more or less straightforward way, as was the case for the sample events analysed in this study. Unfortunately after the loss of STEREO-B and with STEREO-A slowly approaching Earth, the capability to observe stealth CMEs from a viewpoint well-separated from the Sun--Earth line will be lost at least for a while. In the longer term, observations away from the Sun--Earth line, made for example by a STEREO-like, polar, or L4/L5 mission, would help to provide this capability \citep[e.g.,][]{vourlidas2015,lavraud2016,gibson2018,bemporad2021}.

The four events analysed here took place during different stages of the solar cycle, with Event~1 happening at solar minimum, Event~2 and Event~3 close to solar maximum, and Event~4 in the midst of the descending phase of the cycle. Together with the different characteristics of each eruption summarised in Table~\ref{tab:results}, this indicates that stealth CMEs are not restricted to a particular set of source regions or solar activity period. Hence, although the long-standing question on the fundamental nature of stealth CMEs has not been officially answered yet, this study yet emphasises that these events can present as diverse characteristics as ``ordinary'' eruptions. Other methods that may help advance current understanding of stealth CMEs include the study of the coronal environment from which these eruptions originate, as was done by \citet{okane2021a} for Event~2 in this work. The authors concluded that flux emergence and magnetic reconnection episodes were observed in the CME source region prior to eruption, which led to the formation of the structure that later left the Sun as a stealth event, and that a high-altitude null point was revealed by photospheric magnetic field extrapolations of the pre-eruptive configuration. Moreover, images of the solar disc taken closer than 1~AU may more easily reveal the eruptive signatures of stealth CMEs. The Solar Orbiter \citep{muller2020} spacecraft, launched in February 2020 to orbit the Sun as close as ${\sim}0.3$~AU and equipped with an EUV instrument as well as a coronagraph, will possibly be able to provide answers in this regard. Finally, it is worth remarking on the impact of stealth CMEs in the wider context of space weather. These events are occasionally capable of driving large geomagnetic disturbances \citep[e.g.,][]{nitta2017}, so it is important to develop a framework in which they can be fully observed and forecast. One major issue in this sense is that the lack of well-defined low-coronal signatures does not allow for unambiguous analysis of the remote-sensing proxies that are necessary to determine the CME pre-eruptive structure and configuration \citep[e.g.,][]{palmerio2017}. Nevertheless, successfully identifying the source region of a stealth CME represents a first step towards providing more reliable predictions.


\section*{Conflict of Interest Statement}

The authors declare that the research was conducted in the absence of any commercial or financial relationships that could be construed as a potential conflict of interest.

\section*{Author Contributions}

EP and NVN selected the events examined in this study and organised the various aspects of the analysis. EP led the investigation and wrote the draft manuscript. All authors participated in the discussions, read and critically reviewed the manuscript, approved the final version, and agreed to be accountable for all aspects of the work.

\section*{Funding}

EP was supported by the NASA Living With a Star Jack Eddy Postdoctoral Fellowship Program, administered by UCAR’s Cooperative Programs for the Advancement of Earth System Science (CPAESS) under award no. NNX16AK22G. NVN and TM acknowledge support from NASA grant no. NNX17AB73G. NVN was partially supported by the NASA AIA contract no. NNG04EA00C and the NASA STEREO mission under NRL contract no. N00173-02-C-2035. MM and ANZ thank the European Space Agency (ESA) and the Belgian Federal Science Policy Office (BELSPO) for their support in the framework of the PRODEX Programme. JO thanks the STFC for support via funding given in her PHD studentship. IGR was supported by the STEREO mission and NASA program NNH17ZDA001N-LWS.

\section*{Acknowledgements}

The authors are pleased to acknowledge the International Space Science Institute (ISSI) for their support of International Team no. 415, ``Understanding the Origins of Problem Geomagnetic Storms'' (\url{https://www.issibern.ch/teams/geomagstorm/}), from which this work originated.

\section*{Supplemental Data}

The Supplementary Material for this article can be found online at: \url{https://www.frontiersin.org/articles/10.3389/fspas.2021.695966/}


\section*{Data Availability Statement}

All the solar data available in this study can be accessed at the Virtual Solar Observatory (\url{http://sdac.virtualsolar.org/cgi/search}). These data were visualised, processed, and analysed using Python's SunPy package \citep{sunpy2020}, IDL SolarSoft \citep{freeland1998}, and the ESA JHelioviewer software \citep{muller2017}.


\bibliographystyle{frontiersinSCNS_ENG_HUMS}
\bibliography{bibliography}

\begin{thebibliography}{103}
\providecommand{\natexlab}[1]{#1}
\expandafter\ifx\csname urlstyle\endcsname\relax
  \providecommand{\doi}[1]{doi:\discretionary{}{}{}#1}\else
  \providecommand{\doi}{doi:\discretionary{}{}{}\begingroup
  \urlstyle{rm}\Url}\fi
\providecommand{\selectlanguage}[1]{\relax}
\providecommand{\bibAnnoteFile}[1]{%
  \IfFileExists{#1}{\begin{quotation}\noindent\textsc{Key:} #1\\
  \textsc{Annotation:}\ \input{#1}\end{quotation}}{}}
\providecommand{\bibAnnote}[2]{%
  \begin{quotation}\noindent\textsc{Key:} #1\\
  \textsc{Annotation:}\ #2\end{quotation}}

\bibitem[{{Alzate} and {Morgan}(2017)}]{alzate2017}
{Alzate}, N. and {Morgan}, H. (2017).
\newblock {Identification of Low Coronal Sources of
  {\textquotedblleft}Stealth{\textquotedblright} Coronal Mass Ejections Using
  New Image Processing Techniques}.
\newblock \emph{\apj} 840, 103.
\newblock \doi{10.3847/1538-4357/aa6caa}
\bibAnnoteFile{alzate2017}

\bibitem[{{Attrill} et~al.(2007){Attrill}, {Harra}, {van Driel-Gesztelyi}, and
  {D{\'e}moulin}}]{attrill2007}
{Attrill}, G. D.~R., {Harra}, L.~K., {van Driel-Gesztelyi}, L., and
  {D{\'e}moulin}, P. (2007).
\newblock {Coronal ``Wave'': Magnetic Footprint of a Coronal Mass Ejection?}
\newblock \emph{\apjl} 656, L101--L104.
\newblock \doi{10.1086/512854}
\bibAnnoteFile{attrill2007}

\bibitem[{{Attrill} and {Wills-Davey}(2010)}]{attrill2010}
{Attrill}, G.~D.~R. and {Wills-Davey}, M.~J. (2010).
\newblock {Automatic Detection and Extraction of Coronal Dimmings from SDO/AIA
  Data}.
\newblock \emph{\solphys} 262, 461--480.
\newblock \doi{10.1007/s11207-009-9444-4}
\bibAnnoteFile{attrill2010}

\bibitem[{{Bemporad}(2021)}]{bemporad2021}
{Bemporad}, A. (2021).
\newblock {Possible Advantages of a Twin Spacecraft Heliospheric Mission at the
  Sun-Earth Lagrangian Points L4 and L5}.
\newblock \emph{\frass} 8, 627576.
\newblock \doi{10.3389/fspas.2021.627576}
\bibAnnoteFile{bemporad2021}

\bibitem[{{Bemporad} et~al.(2011){Bemporad}, {Mierla}, and
  {Tripathi}}]{bemporad2011}
{Bemporad}, A., {Mierla}, M., and {Tripathi}, D. (2011).
\newblock {Rotation of an erupting filament observed by the STEREO EUVI and
  COR1 instruments}.
\newblock \emph{\aap} 531, A147.
\newblock \doi{10.1051/0004-6361/201016297}
\bibAnnoteFile{bemporad2011}

\bibitem[{{Brueckner} et~al.(1995){Brueckner}, {Howard}, {Koomen}, {Korendyke},
  {Michels}, {Moses} et~al.}]{brueckner1995}
{Brueckner}, G.~E., {Howard}, R.~A., {Koomen}, M.~J., {Korendyke}, C.~M.,
  {Michels}, D.~J., {Moses}, J.~D., et~al. (1995).
\newblock {The Large Angle Spectroscopic Coronagraph (LASCO)}.
\newblock \emph{\solphys} 162, 357--402.
\newblock \doi{10.1007/BF00733434}
\bibAnnoteFile{brueckner1995}

\bibitem[{{Burlaga} et~al.(1982){Burlaga}, {Klein}, {Sheeley}, {Michels},
  {Howard}, {Koomen} et~al.}]{burlaga1982}
{Burlaga}, L.~F., {Klein}, L., {Sheeley}, J., N.~R., {Michels}, D.~J.,
  {Howard}, R.~A., {Koomen}, M.~J., et~al. (1982).
\newblock {A magnetic cloud and a coronal mass ejection}.
\newblock \emph{\grl} 9, 1317--1320.
\newblock \doi{10.1029/GL009i012p01317}
\bibAnnoteFile{burlaga1982}

\bibitem[{{Cane} and {Richardson}(2003)}]{cane2003}
{Cane}, H.~V. and {Richardson}, I.~G. (2003).
\newblock {Interplanetary coronal mass ejections in the near-Earth solar wind
  during 1996-2002}.
\newblock \emph{\jgr} 108, 1156.
\newblock \doi{10.1029/2002JA009817}
\bibAnnoteFile{cane2003}

\bibitem[{{Delaboudini{\`e}re} et~al.(1995){Delaboudini{\`e}re}, {Artzner},
  {Brunaud}, {Gabriel}, {Hochedez}, {Millier} et~al.}]{delaboudiniere1995}
{Delaboudini{\`e}re}, J.~P., {Artzner}, G.~E., {Brunaud}, J., {Gabriel}, A.~H.,
  {Hochedez}, J.~F., {Millier}, F., et~al. (1995).
\newblock {EIT: Extreme-Ultraviolet Imaging Telescope for the SOHO Mission}.
\newblock \emph{\solphys} 162, 291--312.
\newblock \doi{10.1007/BF00733432}
\bibAnnoteFile{delaboudiniere1995}

\bibitem[{{Demastus} et~al.(1973){Demastus}, {Wagner}, and
  {Robinson}}]{demastus1973}
{Demastus}, H.~L., {Wagner}, W.~J., and {Robinson}, R.~D. (1973).
\newblock {Coronal Disturbances. I: Fast Transient Events Observed in the Green
  Coronal Emission Line During the Last Solar Cycle}.
\newblock \emph{\solphys} 31, 449--459.
\newblock \doi{10.1007/BF00152820}
\bibAnnoteFile{demastus1973}

\bibitem[{{D'Huys} et~al.(2014){D'Huys}, {Seaton}, {Poedts}, and
  {Berghmans}}]{dhuys2014}
{D'Huys}, E., {Seaton}, D.~B., {Poedts}, S., and {Berghmans}, D. (2014).
\newblock {Observational Characteristics of Coronal Mass Ejections without
  Low-coronal Signatures}.
\newblock \emph{\apj} 795, 49.
\newblock \doi{10.1088/0004-637X/795/1/49}
\bibAnnoteFile{dhuys2014}

\bibitem[{{Domingo} et~al.(1995){Domingo}, {Fleck}, and {Poland}}]{domingo1995}
{Domingo}, V., {Fleck}, B., and {Poland}, A.~I. (1995).
\newblock {The SOHO Mission: an Overview}.
\newblock \emph{\solphys} 162, 1--37.
\newblock \doi{10.1007/BF00733425}
\bibAnnoteFile{domingo1995}

\bibitem[{{Freeland} and {Handy}(1998)}]{freeland1998}
{Freeland}, S.~L. and {Handy}, B.~N. (1998).
\newblock {Data Analysis with the SolarSoft System}.
\newblock \emph{\solphys} 182, 497--500.
\newblock \doi{10.1023/A:1005038224881}
\bibAnnoteFile{freeland1998}

\bibitem[{{Freiherr von Forstner} et~al.(2021){Freiherr von Forstner},
  {Dumbovi{\'c}}, {M{\"o}stl}, {Guo}, {Papaioannou}, {Elftmann}
  et~al.}]{freiherrvonforstner2021}
{Freiherr von Forstner}, J.~L., {Dumbovi{\'c}}, M., {M{\"o}stl}, C., {Guo}, J.,
  {Papaioannou}, A., {Elftmann}, R., et~al. (2021).
\newblock {Radial Evolution of the April 2020 Stealth Coronal Mass Ejection
  between 0.8 and 1 AU -- A Comparison of Forbush Decreases at Solar Orbiter
  and Earth}.
\newblock \emph{\aap} in press.
\newblock \doi{10.1051/0004-6361/202039848}
\bibAnnoteFile{freiherrvonforstner2021}

\bibitem[{{Gibson} et~al.(2006){Gibson}, {Foster}, {Burkepile}, {de Toma}, and
  {Stanger}}]{gibson2006}
{Gibson}, S.~E., {Foster}, D., {Burkepile}, J., {de Toma}, G., and {Stanger},
  A. (2006).
\newblock {The Calm before the Storm: The Link between Quiescent Cavities and
  Coronal Mass Ejections}.
\newblock \emph{\apj} 641, 590--605.
\newblock \doi{10.1086/500446}
\bibAnnoteFile{gibson2006}

\bibitem[{{Gibson} et~al.(2018){Gibson}, {Vourlidas}, {Hassler}, {Rachmeler},
  {Thompson}, {Newmark} et~al.}]{gibson2018}
{Gibson}, S.~E., {Vourlidas}, A., {Hassler}, D.~M., {Rachmeler}, L.~A.,
  {Thompson}, M.~J., {Newmark}, J., et~al. (2018).
\newblock {Solar Physics from Unconventional Viewpoints}.
\newblock \emph{\frass} 5, 32.
\newblock \doi{10.3389/fspas.2018.00032}
\bibAnnoteFile{gibson2018}

\bibitem[{{Gosling}(1975)}]{gosling1975}
{Gosling}, J.~T. (1975).
\newblock {Large-scale inhomogeneities in the solar wind of solar origin.}
\newblock \emph{rvgsp} 13, 1053--1058.
\newblock \doi{10.1029/RG013i003p01053}
\bibAnnoteFile{gosling1975}

\bibitem[{{Gosling} et~al.(1974){Gosling}, {Hildner}, {MacQueen}, {Munro},
  {Poland}, and {Ross}}]{gosling1974}
{Gosling}, J.~T., {Hildner}, E., {MacQueen}, R.~M., {Munro}, R.~H., {Poland},
  A.~I., and {Ross}, C.~L. (1974).
\newblock {Mass ejections from the Sun: A view from Skylab}.
\newblock \emph{\jgr} 79, 4581.
\newblock \doi{10.1029/JA079i031p04581}
\bibAnnoteFile{gosling1974}

\bibitem[{{Green} et~al.(2007){Green}, {Kliem}, {T{\"o}r{\"o}k}, {van
  Driel-Gesztelyi}, and {Attrill}}]{green2007}
{Green}, L.~M., {Kliem}, B., {T{\"o}r{\"o}k}, T., {van Driel-Gesztelyi}, L.,
  and {Attrill}, G.~D.~R. (2007).
\newblock {Transient Coronal Sigmoids and Rotating Erupting Flux Ropes}.
\newblock \emph{\solphys} 246, 365--391.
\newblock \doi{10.1007/s11207-007-9061-z}
\bibAnnoteFile{green2007}

\bibitem[{{Halain} et~al.(2013){Halain}, {Berghmans}, {Seaton}, {Nicula}, {De
  Groof}, {Mierla} et~al.}]{halain2013}
{Halain}, J.~P., {Berghmans}, D., {Seaton}, D.~B., {Nicula}, B., {De Groof},
  A., {Mierla}, M., et~al. (2013).
\newblock {The SWAP EUV Imaging Telescope. Part II: In-flight Performance and
  Calibration}.
\newblock \emph{\solphys} 286, 67--91.
\newblock \doi{10.1007/s11207-012-0183-6}
\bibAnnoteFile{halain2013}

\bibitem[{{He} et~al.(2018){He}, {Liu}, {Hu}, {Wang}, and {Zhao}}]{he2018}
{He}, W., {Liu}, Y.~D., {Hu}, H., {Wang}, R., and {Zhao}, X. (2018).
\newblock {A Stealth CME Bracketed between Slow and Fast Wind Producing
  Unexpected Geoeffectiveness}.
\newblock \emph{\apj} 860, 78.
\newblock \doi{10.3847/1538-4357/aac381}
\bibAnnoteFile{he2018}

\bibitem[{{Howard} et~al.(1982){Howard}, {Michels}, {Sheeley}, and
  {Koomen}}]{howard1982}
{Howard}, R.~A., {Michels}, D.~J., {Sheeley}, J., N.~R., and {Koomen}, M.~J.
  (1982).
\newblock {The observation of a coronal transient directed at Earth.}
\newblock \emph{\apjl} 263, L101--L104.
\newblock \doi{10.1086/183932}
\bibAnnoteFile{howard1982}

\bibitem[{{Howard} et~al.(2008){Howard}, {Moses}, {Vourlidas}, {Newmark},
  {Socker}, {Plunkett} et~al.}]{howard2008a}
{Howard}, R.~A., {Moses}, J.~D., {Vourlidas}, A., {Newmark}, J.~S., {Socker},
  D.~G., {Plunkett}, S.~P., et~al. (2008).
\newblock {Sun Earth Connection Coronal and Heliospheric Investigation
  (SECCHI)}.
\newblock \emph{\ssr} 136, 67--115.
\newblock \doi{10.1007/s11214-008-9341-4}
\bibAnnoteFile{howard2008a}

\bibitem[{{Howard} and {Harrison}(2013)}]{howard2013}
{Howard}, T.~A. and {Harrison}, R.~A. (2013).
\newblock {Stealth Coronal Mass Ejections: A Perspective}.
\newblock \emph{\solphys} 285, 269--280.
\newblock \doi{10.1007/s11207-012-0217-0}
\bibAnnoteFile{howard2013}

\bibitem[{{Hudson} et~al.(1992){Hudson}, {Acton}, {Hirayama}, and
  {Uchida}}]{hudson1992}
{Hudson}, H.~S., {Acton}, L.~W., {Hirayama}, T., and {Uchida}, Y. (1992).
\newblock {White-Light Flares Observed by YOHKOH}.
\newblock \emph{\pasj} 44, L77--L81
\bibAnnoteFile{hudson1992}

\bibitem[{{Hudson} and {Cliver}(2001)}]{hudson2001}
{Hudson}, H.~S. and {Cliver}, E.~W. (2001).
\newblock {Observing coronal mass ejections without coronagraphs}.
\newblock \emph{\jgr} 106, 25199--25214.
\newblock \doi{10.1029/2000JA904026}
\bibAnnoteFile{hudson2001}

\bibitem[{{Illing} and {Hundhausen}(1985)}]{illing1985}
{Illing}, R.~M.~E. and {Hundhausen}, A.~J. (1985).
\newblock {Observation of a coronal transient from 1.2 to 6 solar radii}.
\newblock \emph{\jgr} 90, 275--282.
\newblock \doi{10.1029/JA090iA01p00275}
\bibAnnoteFile{illing1985}

\bibitem[{{Inhester}(2006)}]{inhester2006}
{Inhester}, B. (2006).
\newblock {Stereoscopy basics for the STEREO mission}.
\newblock \emph{arXiv e-prints} , astro-ph/0612649
\bibAnnoteFile{inhester2006}

\bibitem[{{Isavnin} et~al.(2014){Isavnin}, {Vourlidas}, and
  {Kilpua}}]{isavnin2014}
{Isavnin}, A., {Vourlidas}, A., and {Kilpua}, E.~K.~J. (2014).
\newblock {Three-Dimensional Evolution of Flux-Rope CMEs and Its Relation to
  the Local Orientation of the Heliospheric Current Sheet}.
\newblock \emph{\solphys} 289, 2141--2156.
\newblock \doi{10.1007/s11207-013-0468-4}
\bibAnnoteFile{isavnin2014}

\bibitem[{{Kahler} and {Hudson}(2001)}]{kahler2001}
{Kahler}, S.~W. and {Hudson}, H.~S. (2001).
\newblock {Origin and development of transient coronal holes}.
\newblock \emph{\jgr} 106, 29239--29248.
\newblock \doi{10.1029/2001JA000127}
\bibAnnoteFile{kahler2001}

\bibitem[{{Kaiser} et~al.(2008){Kaiser}, {Kucera}, {Davila}, {St. Cyr},
  {Guhathakurta}, and {Christian}}]{kaiser2008}
{Kaiser}, M.~L., {Kucera}, T.~A., {Davila}, J.~M., {St. Cyr}, O.~C.,
  {Guhathakurta}, M., and {Christian}, E. (2008).
\newblock {The STEREO Mission: An Introduction}.
\newblock \emph{\ssr} 136, 5--16.
\newblock \doi{10.1007/s11214-007-9277-0}
\bibAnnoteFile{kaiser2008}

\bibitem[{{Kay} and {Opher}(2015)}]{kay2015b}
{Kay}, C. and {Opher}, M. (2015).
\newblock {The Heliocentric Distance where the Deflections and Rotations of
  Solar Coronal Mass Ejections Occur}.
\newblock \emph{\apjl} 811, L36.
\newblock \doi{10.1088/2041-8205/811/2/L36}
\bibAnnoteFile{kay2015b}

\bibitem[{{Kay} et~al.(2015){Kay}, {Opher}, and {Evans}}]{kay2015a}
{Kay}, C., {Opher}, M., and {Evans}, R.~M. (2015).
\newblock {Global Trends of CME Deflections Based on CME and Solar Parameters}.
\newblock \emph{\apj} 805, 168.
\newblock \doi{10.1088/0004-637X/805/2/168}
\bibAnnoteFile{kay2015a}

\bibitem[{{Kilpua} et~al.(2014){Kilpua}, {Mierla}, {Zhukov}, {Rodriguez},
  {Vourlidas}, and {Wood}}]{kilpua2014}
{Kilpua}, E.~K.~J., {Mierla}, M., {Zhukov}, A.~N., {Rodriguez}, L.,
  {Vourlidas}, A., and {Wood}, B. (2014).
\newblock {Solar Sources of Interplanetary Coronal Mass Ejections During the
  Solar Cycle 23/24 Minimum}.
\newblock \emph{\solphys} 289, 3773--3797.
\newblock \doi{10.1007/s11207-014-0552-4}
\bibAnnoteFile{kilpua2014}

\bibitem[{{Koomen} et~al.(1975){Koomen}, {Detwiler}, {Brueckner}, {Cooper}, and
  {Tousey}}]{koomen1975}
{Koomen}, M.~J., {Detwiler}, C.~R., {Brueckner}, G.~E., {Cooper}, H.~W., and
  {Tousey}, R. (1975).
\newblock {White light coronagraph in OSO-7}.
\newblock \emph{\ao} 14, 743--751.
\newblock \doi{10.1364/AO.14.000743}
\bibAnnoteFile{koomen1975}

\bibitem[{{Lavraud} et~al.(2016){Lavraud}, {Liu}, {Segura}, {He}, {Qin},
  {Temmer} et~al.}]{lavraud2016}
{Lavraud}, B., {Liu}, Y., {Segura}, K., {He}, J., {Qin}, G., {Temmer}, M.,
  et~al. (2016).
\newblock {A small mission concept to the Sun-Earth Lagrangian L5 point for
  innovative solar, heliospheric and space weather science}.
\newblock \emph{\jastp} 146, 171--185.
\newblock \doi{10.1016/j.jastp.2016.06.004}
\bibAnnoteFile{lavraud2016}

\bibitem[{{Lemen} et~al.(2012){Lemen}, {Title}, {Akin}, {Boerner}, {Chou},
  {Drake} et~al.}]{lemen2012}
{Lemen}, J.~R., {Title}, A.~M., {Akin}, D.~J., {Boerner}, P.~F., {Chou}, C.,
  {Drake}, J.~F., et~al. (2012).
\newblock {The Atmospheric Imaging Assembly (AIA) on the Solar Dynamics
  Observatory (SDO)}.
\newblock \emph{\solphys} 275, 17--40.
\newblock \doi{10.1007/s11207-011-9776-8}
\bibAnnoteFile{lemen2012}

\bibitem[{{Liewer} et~al.(2015){Liewer}, {Panasenco}, {Vourlidas}, and
  {Colaninno}}]{liewer2015}
{Liewer}, P., {Panasenco}, O., {Vourlidas}, A., and {Colaninno}, R. (2015).
\newblock {Observations and Analysis of the Non-Radial Propagation of Coronal
  Mass Ejections Near the Sun}.
\newblock \emph{\solphys} 290, 3343--3364.
\newblock \doi{10.1007/s11207-015-0794-9}
\bibAnnoteFile{liewer2015}

\bibitem[{{Liewer} et~al.(2011){Liewer}, {Hall}, {Howard}, {De Jong},
  {Thompson}, and {Thernisien}}]{liewer2011}
{Liewer}, P.~C., {Hall}, J.~R., {Howard}, R.~A., {De Jong}, E.~M., {Thompson},
  W.~T., and {Thernisien}, A. (2011).
\newblock {Stereoscopic analysis of STEREO/SECCHI data for CME trajectory
  determination}.
\newblock \emph{\jastp} 73, 1173--1186.
\newblock \doi{10.1016/j.jastp.2010.09.004}
\bibAnnoteFile{liewer2011}

\bibitem[{{Liewer} et~al.(2021){Liewer}, {Qiu}, {Vourlidas}, {Hall}, and
  {Penteado}}]{liewer2021}
{Liewer}, P.~C., {Qiu}, J., {Vourlidas}, A., {Hall}, J.~R., and {Penteado}, P.
  (2021).
\newblock {Evolution of a streamer-blowout CME as observed by imagers on Parker
  Solar Probe and the Solar Terrestrial Relations Observatory}.
\newblock \emph{\aap} 650, A32.
\newblock \doi{10.1051/0004-6361/202039641}
\bibAnnoteFile{liewer2021}

\bibitem[{{Lynch} et~al.(2010){Lynch}, {Li}, {Thernisien}, {Robbrecht},
  {Fisher}, {Luhmann} et~al.}]{lynch2010}
{Lynch}, B.~J., {Li}, Y., {Thernisien}, A.~F.~R., {Robbrecht}, E., {Fisher},
  G.~H., {Luhmann}, J.~G., et~al. (2010).
\newblock {Sun to 1 AU propagation and evolution of a slow streamer-blowout
  coronal mass ejection}.
\newblock \emph{\jgr} 115, A07106.
\newblock \doi{10.1029/2009JA015099}
\bibAnnoteFile{lynch2010}

\bibitem[{{Lynch} et~al.(2016){Lynch}, {Masson}, {Li}, {DeVore}, {Luhmann},
  {Antiochos} et~al.}]{lynch2016}
{Lynch}, B.~J., {Masson}, S., {Li}, Y., {DeVore}, C.~R., {Luhmann}, J.~G.,
  {Antiochos}, S.~K., et~al. (2016).
\newblock {A model for stealth coronal mass ejections}.
\newblock \emph{\jgra} 121, 10,677--10,697.
\newblock \doi{10.1002/2016JA023432}
\bibAnnoteFile{lynch2016}

\bibitem[{{Ma} et~al.(2010){Ma}, {Attrill}, {Golub}, and {Lin}}]{ma2010}
{Ma}, S., {Attrill}, G.~D.~R., {Golub}, L., and {Lin}, J. (2010).
\newblock {Statistical Study of Coronal Mass Ejections With and Without
  Distinct Low Coronal Signatures}.
\newblock \emph{\apj} 722, 289--301.
\newblock \doi{10.1088/0004-637X/722/1/289}
\bibAnnoteFile{ma2010}

\bibitem[{{MacQueen} et~al.(1974){MacQueen}, {Eddy}, {Gosling}, {Hildner},
  {Munro}, {Newkirk} et~al.}]{macqueen1974}
{MacQueen}, R.~M., {Eddy}, J.~A., {Gosling}, J.~T., {Hildner}, E., {Munro},
  R.~H., {Newkirk}, J., G.~A., et~al. (1974).
\newblock {The Outer Solar Corona as Observed from Skylab: Preliminary
  Results}.
\newblock \emph{\apjl} 187, L85.
\newblock \doi{10.1086/181402}
\bibAnnoteFile{macqueen1974}

\bibitem[{{Martin}(1979)}]{martin1979}
{Martin}, S.~F. (1979).
\newblock {Study of the Post-Flare Loops on 1973JULY29 - Part Three - Dynamics
  of the H$\alpha$ Loops}.
\newblock \emph{\solphys} 64, 165--176.
\newblock \doi{10.1007/BF00151125}
\bibAnnoteFile{martin1979}

\bibitem[{{McAllister} et~al.(1996){McAllister}, {Dryer}, {McIntosh}, {Singer},
  and {Weiss}}]{mcallister1996}
{McAllister}, A.~H., {Dryer}, M., {McIntosh}, P., {Singer}, H., and {Weiss}, L.
  (1996).
\newblock {A large polar crown coronal mass ejection and a ``problem''
  geomagnetic storm: April 14-23, 1994}.
\newblock \emph{\jgr} 101, 13497--13516.
\newblock \doi{10.1029/96JA00510}
\bibAnnoteFile{mcallister1996}

\bibitem[{{Michels} et~al.(1980){Michels}, {Howard}, {Koomen}, and
  {Sheeley}}]{michels1980}
{Michels}, D.~J., {Howard}, R.~A., {Koomen}, M.~J., and {Sheeley}, J., N.~R.
  (1980).
\newblock {Satellite observations of the outer corona near sunspot maximum}.
\newblock In \emph{Radio Physics of the Sun, IAS Symposium}, eds. M.~R. {Kundu}
  and T.~E. {Gergely}. vol.~86, 439--442.
\newblock \doi{10.1017/S0074180900037190}
\bibAnnoteFile{michels1980}

\bibitem[{{Mierla} et~al.(2008){Mierla}, {Davila}, {Thompson}, {Inhester},
  {Srivastava}, {Kramar} et~al.}]{mierla2008}
{Mierla}, M., {Davila}, J., {Thompson}, W., {Inhester}, B., {Srivastava}, N.,
  {Kramar}, M., et~al. (2008).
\newblock {A Quick Method for Estimating the Propagation Direction of Coronal
  Mass Ejections Using STEREO-COR1 Images}.
\newblock \emph{\solphys} 252, 385--396.
\newblock \doi{10.1007/s11207-008-9267-8}
\bibAnnoteFile{mierla2008}

\bibitem[{{Mierla} et~al.(2010){Mierla}, {Inhester}, {Antunes}, {Boursier},
  {Byrne}, {Colaninno} et~al.}]{mierla2010}
{Mierla}, M., {Inhester}, B., {Antunes}, A., {Boursier}, Y., {Byrne}, J.~P.,
  {Colaninno}, R., et~al. (2010).
\newblock {On the 3-D reconstruction of Coronal Mass Ejections using
  coronagraph data}.
\newblock \emph{\angeo} 28, 203--215.
\newblock \doi{10.5194/angeo-28-203-2010}
\bibAnnoteFile{mierla2010}

\bibitem[{{Mierla} et~al.(2009){Mierla}, {Inhester}, {Marqu{\'e}}, {Rodriguez},
  {Gissot}, {Zhukov} et~al.}]{mierla2009}
{Mierla}, M., {Inhester}, B., {Marqu{\'e}}, C., {Rodriguez}, L., {Gissot}, S.,
  {Zhukov}, A.~N., et~al. (2009).
\newblock {On 3D Reconstruction of Coronal Mass Ejections: I. Method
  Description and Application to SECCHI-COR Data}.
\newblock \emph{\solphys} 259, 123--141.
\newblock \doi{10.1007/s11207-009-9416-8}
\bibAnnoteFile{mierla2009}

\bibitem[{{Morgan} and {Druckm{\"u}ller}(2014)}]{morgan2014}
{Morgan}, H. and {Druckm{\"u}ller}, M. (2014).
\newblock {Multi-Scale Gaussian Normalization for Solar Image Processing}.
\newblock \emph{\solphys} 289, 2945--2955.
\newblock \doi{10.1007/s11207-014-0523-9}
\bibAnnoteFile{morgan2014}

\bibitem[{{M{\"o}stl} et~al.(2009){M{\"o}stl}, {Farrugia}, {Temmer},
  {Miklenic}, {Veronig}, {Galvin} et~al.}]{mostl2009}
{M{\"o}stl}, C., {Farrugia}, C.~J., {Temmer}, M., {Miklenic}, C., {Veronig},
  A.~M., {Galvin}, A.~B., et~al. (2009).
\newblock {Linking Remote Imagery of a Coronal Mass Ejection to Its In Situ
  Signatures at 1 AU}.
\newblock \emph{\apjl} 705, L180--L185.
\newblock \doi{10.1088/0004-637X/705/2/L180}
\bibAnnoteFile{mostl2009}

\bibitem[{{M{\"u}ller} et~al.(2017){M{\"u}ller}, {Nicula}, {Felix},
  {Verstringe}, {Bourgoignie}, {Csillaghy} et~al.}]{muller2017}
{M{\"u}ller}, D., {Nicula}, B., {Felix}, S., {Verstringe}, F., {Bourgoignie},
  B., {Csillaghy}, A., et~al. (2017).
\newblock {JHelioviewer. Time-dependent 3D visualisation of solar and
  heliospheric data}.
\newblock \emph{\aap} 606, A10.
\newblock \doi{10.1051/0004-6361/201730893}
\bibAnnoteFile{muller2017}

\bibitem[{{M{\"u}ller} et~al.(2020){M{\"u}ller}, {St. Cyr}, {Zouganelis},
  {Gilbert}, {Marsden}, {Nieves-Chinchilla} et~al.}]{muller2020}
{M{\"u}ller}, D., {St. Cyr}, O.~C., {Zouganelis}, I., {Gilbert}, H.~R.,
  {Marsden}, R., {Nieves-Chinchilla}, T., et~al. (2020).
\newblock {The Solar Orbiter mission. Science overview}.
\newblock \emph{\aap} 642, A1.
\newblock \doi{10.1051/0004-6361/202038467}
\bibAnnoteFile{muller2020}

\bibitem[{{Munro} et~al.(1979){Munro}, {Gosling}, {Hildner}, {MacQueen},
  {Poland}, and {Ross}}]{munro1979}
{Munro}, R.~H., {Gosling}, J.~T., {Hildner}, E., {MacQueen}, R.~M., {Poland},
  A.~I., and {Ross}, C.~L. (1979).
\newblock {The association of coronal mass ejection transients with other forms
  of solar activity.}
\newblock \emph{\solphys} 61, 201--215.
\newblock \doi{10.1007/BF00155456}
\bibAnnoteFile{munro1979}

\bibitem[{{Nieves-Chinchilla} et~al.(2011){Nieves-Chinchilla},
  {G{\'o}mez-Herrero}, {Vi{\~n}as}, {Malandraki}, {Dresing}, {Hidalgo}
  et~al.}]{nieveschinchilla2011}
{Nieves-Chinchilla}, T., {G{\'o}mez-Herrero}, R., {Vi{\~n}as}, A.~F.,
  {Malandraki}, O., {Dresing}, N., {Hidalgo}, M.~A., et~al. (2011).
\newblock {Analysis and study of the in situ observation of the June 1st 2008
  CME by STEREO}.
\newblock \emph{\jastp} 73, 1348--1360.
\newblock \doi{10.1016/j.jastp.2010.09.026}
\bibAnnoteFile{nieveschinchilla2011}

\bibitem[{{Nieves-Chinchilla} et~al.(2013){Nieves-Chinchilla}, {Vourlidas},
  {Stenborg}, {Savani}, {Koval}, {Szabo} et~al.}]{nieveschinchilla2013}
{Nieves-Chinchilla}, T., {Vourlidas}, A., {Stenborg}, G., {Savani}, N.~P.,
  {Koval}, A., {Szabo}, A., et~al. (2013).
\newblock {Inner Heliospheric Evolution of a ``Stealth'' CME Derived from
  Multi-view Imaging and Multipoint in Situ observations. I. Propagation to 1
  AU}.
\newblock \emph{\apj} 779, 55.
\newblock \doi{10.1088/0004-637X/779/1/55}
\bibAnnoteFile{nieveschinchilla2013}

\bibitem[{{Nitta} and {Mulligan}(2017)}]{nitta2017}
{Nitta}, N.~V. and {Mulligan}, T. (2017).
\newblock {Earth-Affecting Coronal Mass Ejections Without Obvious Low Coronal
  Signatures}.
\newblock \emph{\solphys} 292, 125.
\newblock \doi{10.1007/s11207-017-1147-7}
\bibAnnoteFile{nitta2017}

\bibitem[{{Ogawara} et~al.(1991){Ogawara}, {Takano}, {Kato}, {Kosugi},
  {Tsuneta}, {Watanabe} et~al.}]{ogawara1991}
{Ogawara}, Y., {Takano}, T., {Kato}, T., {Kosugi}, T., {Tsuneta}, S.,
  {Watanabe}, T., et~al. (1991).
\newblock {The SOLAR-A Mission: An Overview}.
\newblock \emph{\solphys} 136, 1--16.
\newblock \doi{10.1007/BF00151692}
\bibAnnoteFile{ogawara1991}

\bibitem[{{O'Kane} et~al.(2019){O'Kane}, {Green}, {Long}, and
  {Reid}}]{okane2019}
{O'Kane}, J., {Green}, L., {Long}, D.~M., and {Reid}, H. (2019).
\newblock {Stealth Coronal Mass Ejections from Active Regions}.
\newblock \emph{\apj} 882, 85.
\newblock \doi{10.3847/1538-4357/ab371b}
\bibAnnoteFile{okane2019}

\bibitem[{{O'Kane} et~al.(2021{\natexlab{a}}){O'Kane}, {Green}, {Davies},
  {M{\"o}stl}, {Hinterreiter}, {von Forstner} et~al.}]{okane2021b}
{O'Kane}, J., {Green}, L.~M., {Davies}, E.~E., {M{\"o}stl}, C., {Hinterreiter},
  J., {von Forstner}, J. L.~F., et~al. (2021{\natexlab{a}}).
\newblock {Solar origins of a strong stealth CME detected by Solar Orbiter}.
\newblock \emph{\aap} in press.
\newblock \doi{10.1051/0004-6361/202140622}
\bibAnnoteFile{okane2021b}

\bibitem[{{O'Kane} et~al.(2021{\natexlab{b}}){O'Kane}, {Mac Cormack},
  {Mandrini}, {D{\'e}moulin}, {Green}, {Long} et~al.}]{okane2021a}
{O'Kane}, J., {Mac Cormack}, C., {Mandrini}, C.~H., {D{\'e}moulin}, P.,
  {Green}, L.~M., {Long}, D.~M., et~al. (2021{\natexlab{b}}).
\newblock {The Magnetic Environment of a Stealth Coronal Mass Ejection}.
\newblock \emph{\apj} 908, 89.
\newblock \doi{10.3847/1538-4357/abd2bf}
\bibAnnoteFile{okane2021a}

\bibitem[{{Palmerio} et~al.(2017){Palmerio}, {Kilpua}, {James}, {Green},
  {Pomoell}, {Isavnin} et~al.}]{palmerio2017}
{Palmerio}, E., {Kilpua}, E.~K.~J., {James}, A.~W., {Green}, L.~M., {Pomoell},
  J., {Isavnin}, A., et~al. (2017).
\newblock {Determining the Intrinsic CME Flux Rope Type Using Remote-sensing
  Solar Disk Observations}.
\newblock \emph{\solphys} 292, 39.
\newblock \doi{10.1007/s11207-017-1063-x}
\bibAnnoteFile{palmerio2017}

\bibitem[{{Palmerio} et~al.(2021){Palmerio}, {Kilpua}, {Witasse}, {Barnes},
  {S{\'a}nchez-Cano}, {Weiss} et~al.}]{palmerio2021}
{Palmerio}, E., {Kilpua}, E. K.~J., {Witasse}, O., {Barnes}, D.,
  {S{\'a}nchez-Cano}, B., {Weiss}, A.~J., et~al. (2021).
\newblock {CME Magnetic Structure and IMF Preconditioning Affecting SEP
  Transport}.
\newblock \emph{\spwea} 19, e2020SW002654.
\newblock \doi{10.1029/2020SW002654}
\bibAnnoteFile{palmerio2021}

\bibitem[{{Panasenco} et~al.(2013){Panasenco}, {Martin}, {Velli}, and
  {Vourlidas}}]{panasenco2013}
{Panasenco}, O., {Martin}, S.~F., {Velli}, M., and {Vourlidas}, A. (2013).
\newblock {Origins of Rolling, Twisting, and Non-radial Propagation of Eruptive
  Solar Events}.
\newblock \emph{\solphys} 287, 391--413.
\newblock \doi{10.1007/s11207-012-0194-3}
\bibAnnoteFile{panasenco2013}

\bibitem[{{Pesnell} et~al.(2012){Pesnell}, {Thompson}, and
  {Chamberlin}}]{pesnell2012}
{Pesnell}, W.~D., {Thompson}, B.~J., and {Chamberlin}, P.~C. (2012).
\newblock {The Solar Dynamics Observatory (SDO)}.
\newblock \emph{\solphys} 275, 3--15.
\newblock \doi{10.1007/s11207-011-9841-3}
\bibAnnoteFile{pesnell2012}

\bibitem[{{Pevtsov} et~al.(2012){Pevtsov}, {Panasenco}, and
  {Martin}}]{pevtsov2012}
{Pevtsov}, A.~A., {Panasenco}, O., and {Martin}, S.~F. (2012).
\newblock {Coronal Mass Ejections from Magnetic Systems Encompassing Filament
  Channels Without Filaments}.
\newblock \emph{\solphys} 277, 185--201.
\newblock \doi{10.1007/s11207-011-9881-8}
\bibAnnoteFile{pevtsov2012}

\bibitem[{{Richardson} and {Cane}(2010)}]{richardson2010}
{Richardson}, I.~G. and {Cane}, H.~V. (2010).
\newblock {Near-Earth Interplanetary Coronal Mass Ejections During Solar Cycle
  23 (1996 - 2009): Catalog and Summary of Properties}.
\newblock \emph{\solphys} 264, 189--237.
\newblock \doi{10.1007/s11207-010-9568-6}
\bibAnnoteFile{richardson2010}

\bibitem[{{Robbrecht} et~al.(2009){Robbrecht}, {Patsourakos}, and
  {Vourlidas}}]{robbrecht2009}
{Robbrecht}, E., {Patsourakos}, S., and {Vourlidas}, A. (2009).
\newblock {No Trace Left Behind: STEREO Observation of a Coronal Mass Ejection
  Without Low Coronal Signatures}.
\newblock \emph{\apj} 701, 283--291.
\newblock \doi{10.1088/0004-637X/701/1/283}
\bibAnnoteFile{robbrecht2009}

\bibitem[{{Rust} and {Bar}(1973)}]{rust1973}
{Rust}, D.~M. and {Bar}, V. (1973).
\newblock {Magnetic fields, loop prominences and the great flares of August,
  1972}.
\newblock \emph{\solphys} 33, 445--459.
\newblock \doi{10.1007/BF00152432}
\bibAnnoteFile{rust1973}

\bibitem[{{Rust} and {Kumar}(1996)}]{rust1996}
{Rust}, D.~M. and {Kumar}, A. (1996).
\newblock {Evidence for Helically Kinked Magnetic Flux Ropes in Solar
  Eruptions}.
\newblock \emph{\apjl} 464, L199.
\newblock \doi{10.1086/310118}
\bibAnnoteFile{rust1996}

\bibitem[{{Rust} et~al.(1975){Rust}, {Nakagawa}, and {Neupert}}]{rust1975}
{Rust}, D.~M., {Nakagawa}, Y., and {Neupert}, W.~M. (1975).
\newblock {EUV Emission, Filament Activation and Magnetic Fields in a Slow-Rise
  Flare}.
\newblock \emph{\solphys} 41, 397--414.
\newblock \doi{10.1007/BF00154077}
\bibAnnoteFile{rust1975}

\bibitem[{{Rust} and {Webb}(1977)}]{rust1977}
{Rust}, D.~M. and {Webb}, D.~F. (1977).
\newblock {Soft X-ray observations of large-scale coronal active region
  brightenings.}
\newblock \emph{\solphys} 54, 403--417.
\newblock \doi{10.1007/BF00159932}
\bibAnnoteFile{rust1977}

\bibitem[{{Santandrea} et~al.(2013){Santandrea}, {Gantois}, {Strauch},
  {Teston}, {Tilmans}, {Baijot} et~al.}]{santandrea2013}
{Santandrea}, S., {Gantois}, K., {Strauch}, K., {Teston}, F., {Tilmans}, E.,
  {Baijot}, C., et~al. (2013).
\newblock {PROBA2: Mission and Spacecraft Overview}.
\newblock \emph{\solphys} 286, 5--19.
\newblock \doi{10.1007/s11207-013-0289-5}
\bibAnnoteFile{santandrea2013}

\bibitem[{{Schmidt} et~al.(2016){Schmidt}, {Cairns}, {Xie}, {St. Cyr}, and
  {Gopalswamy}}]{schmidt2016}
{Schmidt}, J.~M., {Cairns}, I.~H., {Xie}, H., {St. Cyr}, O.~C., and
  {Gopalswamy}, N. (2016).
\newblock {CME flux rope and shock identifications and locations: Comparison of
  white light data, Graduated Cylindrical Shell model, and MHD simulations}.
\newblock \emph{\jgra} 121, 1886--1906.
\newblock \doi{10.1002/2015JA021805}
\bibAnnoteFile{schmidt2016}

\bibitem[{{Schwenn} et~al.(2005){Schwenn}, {dal Lago}, {Huttunen}, and
  {Gonzalez}}]{schwenn2005}
{Schwenn}, R., {dal Lago}, A., {Huttunen}, E., and {Gonzalez}, W.~D. (2005).
\newblock {The association of coronal mass ejections with their effects near
  the Earth}.
\newblock \emph{\angeo} 23, 1033--1059.
\newblock \doi{10.5194/angeo-23-1033-2005}
\bibAnnoteFile{schwenn2005}

\bibitem[{{Seaton} et~al.(2013){Seaton}, {Berghmans}, {Nicula}, {Halain}, {De
  Groof}, {Thibert} et~al.}]{seaton2013}
{Seaton}, D.~B., {Berghmans}, D., {Nicula}, B., {Halain}, J.~P., {De Groof},
  A., {Thibert}, T., et~al. (2013).
\newblock {The SWAP EUV Imaging Telescope Part I: Instrument Overview and
  Pre-Flight Testing}.
\newblock \emph{\solphys} 286, 43--65.
\newblock \doi{10.1007/s11207-012-0114-6}
\bibAnnoteFile{seaton2013}

\bibitem[{{Sheeley} et~al.(1975){Sheeley}, {Bohlin}, {Brueckner}, {Purcell},
  {Scherrer}, {Tousey} et~al.}]{sheeley1975}
{Sheeley}, J., N.~R., {Bohlin}, J.~D., {Brueckner}, G.~E., {Purcell}, J.~D.,
  {Scherrer}, V.~E., {Tousey}, R., et~al. (1975).
\newblock {Coronal Changes Associated with a Disappearing Filament}.
\newblock \emph{\solphys} 45, 377--392.
\newblock \doi{10.1007/BF00158457}
\bibAnnoteFile{sheeley1975}

\bibitem[{{Shensa}(1992)}]{shensa1992}
{Shensa}, M.~J. (1992).
\newblock {The discrete wavelet transform: wedding the a trous and Mallat
  algorithms}.
\newblock \emph{\itsp} 40, 2464--2482.
\newblock \doi{10.1109/78.157290}
\bibAnnoteFile{shensa1992}

\bibitem[{{Shi} et~al.(2015){Shi}, {Wang}, {Wan}, {Cheng}, {Ding}, and
  {Zhang}}]{shi2015}
{Shi}, T., {Wang}, Y., {Wan}, L., {Cheng}, X., {Ding}, M., and {Zhang}, J.
  (2015).
\newblock {Predicting the Arrival Time of Coronal Mass Ejections with the
  Graduated Cylindrical Shell and Drag Force Model}.
\newblock \emph{\apj} 806, 271.
\newblock \doi{10.1088/0004-637X/806/2/271}
\bibAnnoteFile{shi2015}

\bibitem[{{Srivastava} et~al.(2009){Srivastava}, {Inhester}, {Mierla}, and
  {Podlipnik}}]{srivastava2009}
{Srivastava}, N., {Inhester}, B., {Mierla}, M., and {Podlipnik}, B. (2009).
\newblock {3D Reconstruction of the Leading Edge of the 20 May 2007 Partial
  Halo CME}.
\newblock \emph{\solphys} 259, 213--225.
\newblock \doi{10.1007/s11207-009-9423-9}
\bibAnnoteFile{srivastava2009}

\bibitem[{{Stenborg} and {Cobelli}(2003)}]{stenborg2003}
{Stenborg}, G. and {Cobelli}, P.~J. (2003).
\newblock {A wavelet packets equalization technique to reveal the multiple
  spatial-scale nature of coronal structures}.
\newblock \emph{\aap} 398, 1185--1193.
\newblock \doi{10.1051/0004-6361:20021687}
\bibAnnoteFile{stenborg2003}

\bibitem[{{Stenborg} et~al.(2008){Stenborg}, {Vourlidas}, and
  {Howard}}]{stenborg2008}
{Stenborg}, G., {Vourlidas}, A., and {Howard}, R.~A. (2008).
\newblock {A Fresh View of the Extreme-Ultraviolet Corona from the Application
  of a New Image-Processing Technique}.
\newblock \emph{\apj} 674, 1201--1206.
\newblock \doi{10.1086/525556}
\bibAnnoteFile{stenborg2008}

\bibitem[{{SunPy Community} et~al.(2020){SunPy Community}, {Barnes}, {Bobra},
  {Christe}, {Freij}, {Hayes} et~al.}]{sunpy2020}
{SunPy Community}, {Barnes}, W.~T., {Bobra}, M.~G., {Christe}, S.~D., {Freij},
  N., {Hayes}, L.~A., et~al. (2020).
\newblock {The SunPy Project: Open Source Development and Status of the Version
  1.0 Core Package}.
\newblock \emph{\apj} 890, 68.
\newblock \doi{10.3847/1538-4357/ab4f7a}
\bibAnnoteFile{sunpy2020}

\bibitem[{{Talpeanu} et~al.(2020){Talpeanu}, {Chan{\'e}}, {Poedts}, {D'Huys},
  {Mierla}, {Roussev} et~al.}]{talpeanu2020}
{Talpeanu}, D.~C., {Chan{\'e}}, E., {Poedts}, S., {D'Huys}, E., {Mierla}, M.,
  {Roussev}, I., et~al. (2020).
\newblock {Numerical simulations of shear-induced consecutive coronal mass
  ejections}.
\newblock \emph{\aap} 637, A77.
\newblock \doi{10.1051/0004-6361/202037477}
\bibAnnoteFile{talpeanu2020}

\bibitem[{{Thernisien}(2011)}]{thernisien2011}
{Thernisien}, A. (2011).
\newblock {Implementation of the Graduated Cylindrical Shell Model for the
  Three-dimensional Reconstruction of Coronal Mass Ejections}.
\newblock \emph{\apjs} 194, 33.
\newblock \doi{10.1088/0067-0049/194/2/33}
\bibAnnoteFile{thernisien2011}

\bibitem[{{Thernisien} et~al.(2009){Thernisien}, {Vourlidas}, and
  {Howard}}]{thernisien2009}
{Thernisien}, A., {Vourlidas}, A., and {Howard}, R.~A. (2009).
\newblock {Forward Modeling of Coronal Mass Ejections Using STEREO/SECCHI
  Data}.
\newblock \emph{\solphys} 256, 111--130.
\newblock \doi{10.1007/s11207-009-9346-5}
\bibAnnoteFile{thernisien2009}

\bibitem[{{Thernisien} et~al.(2006){Thernisien}, {Howard}, and
  {Vourlidas}}]{thernisien2006}
{Thernisien}, A.~F.~R., {Howard}, R.~A., and {Vourlidas}, A. (2006).
\newblock {Modeling of Flux Rope Coronal Mass Ejections}.
\newblock \emph{\apj} 652, 763--773.
\newblock \doi{10.1086/508254}
\bibAnnoteFile{thernisien2006}

\bibitem[{{Thompson} et~al.(2000){Thompson}, {Cliver}, {Nitta}, {Delann{\'e}e},
  and {Delaboudini{\`e}re}}]{thompson2000}
{Thompson}, B.~J., {Cliver}, E.~W., {Nitta}, N., {Delann{\'e}e}, C., and
  {Delaboudini{\`e}re}, J.-P. (2000).
\newblock {Coronal dimmings and energetic CMEs in April-May 1998}.
\newblock \emph{\grl} 27, 1431--1434.
\newblock \doi{10.1029/1999GL003668}
\bibAnnoteFile{thompson2000}

\bibitem[{{Thompson} et~al.(1998){Thompson}, {Plunkett}, {Gurman}, {Newmark},
  {St. Cyr}, and {Michels}}]{thompson1998}
{Thompson}, B.~J., {Plunkett}, S.~P., {Gurman}, J.~B., {Newmark}, J.~S., {St.
  Cyr}, O.~C., and {Michels}, D.~J. (1998).
\newblock {SOHO/EIT observations of an Earth-directed coronal mass ejection on
  May 12, 1997}.
\newblock \emph{\grl} 25, 2465--2468.
\newblock \doi{10.1029/98GL50429}
\bibAnnoteFile{thompson1998}

\bibitem[{{Thompson}(2009)}]{thompson2009}
{Thompson}, W.~T. (2009).
\newblock {3D triangulation of a Sun-grazing comet}.
\newblock \emph{\icarus} 200, 351--357.
\newblock \doi{10.1016/j.icarus.2008.12.011}
\bibAnnoteFile{thompson2009}

\bibitem[{{Thompson} et~al.(2012){Thompson}, {Kliem}, and
  {T{\"o}r{\"o}k}}]{thompson2012}
{Thompson}, W.~T., {Kliem}, B., and {T{\"o}r{\"o}k}, T. (2012).
\newblock {3D Reconstruction of a Rotating Erupting Prominence}.
\newblock \emph{\solphys} 276, 241--259.
\newblock \doi{10.1007/s11207-011-9868-5}
\bibAnnoteFile{thompson2012}

\bibitem[{{Tousey}(1973)}]{tousey1973}
{Tousey}, R. (1973).
\newblock {The solar corona}.
\newblock In \emph{Space Research Conference}. vol.~2, 713--730
\bibAnnoteFile{tousey1973}

\bibitem[{{Tousey}(1977)}]{tousey1977}
{Tousey}, R. (1977).
\newblock {Apollo-Telescope Mount of Skylab: An overview}.
\newblock \emph{\ao} 16, 825--836.
\newblock \doi{10.1364/AO.16.000825}
\bibAnnoteFile{tousey1977}

\bibitem[{{Tripathi} et~al.(2004){Tripathi}, {Bothmer}, and
  {Cremades}}]{tripathi2004}
{Tripathi}, D., {Bothmer}, V., and {Cremades}, H. (2004).
\newblock {The basic characteristics of EUV post-eruptive arcades and their
  role as tracers of coronal mass ejection source regions}.
\newblock \emph{\aap} 422, 337--349.
\newblock \doi{10.1051/0004-6361:20035815}
\bibAnnoteFile{tripathi2004}

\bibitem[{{Tsuneta} et~al.(1991){Tsuneta}, {Acton}, {Bruner}, {Lemen}, {Brown},
  {Caravalho} et~al.}]{tsuneta1991}
{Tsuneta}, S., {Acton}, L., {Bruner}, M., {Lemen}, J., {Brown}, W.,
  {Caravalho}, R., et~al. (1991).
\newblock {The Soft X-ray Telescope for the SOLAR-A mission}.
\newblock \emph{\solphys} 136, 37--67.
\newblock \doi{10.1007/BF00151694}
\bibAnnoteFile{tsuneta1991}

\bibitem[{{Vourlidas}(2015)}]{vourlidas2015}
{Vourlidas}, A. (2015).
\newblock {Mission to the Sun-Earth L$_{5}$ Lagrangian Point: An Optimal
  Platform for Space Weather Research}.
\newblock \emph{\spwea} 13, 197--201.
\newblock \doi{10.1002/2015SW001173}
\bibAnnoteFile{vourlidas2015}

\bibitem[{{Vourlidas} et~al.(2011){Vourlidas}, {Colaninno},
  {Nieves-Chinchilla}, and {Stenborg}}]{vourlidas2011}
{Vourlidas}, A., {Colaninno}, R., {Nieves-Chinchilla}, T., and {Stenborg}, G.
  (2011).
\newblock {The First Observation of a Rapidly Rotating Coronal Mass Ejection in
  the Middle Corona}.
\newblock \emph{\apjl} 733, L23.
\newblock \doi{10.1088/2041-8205/733/2/L23}
\bibAnnoteFile{vourlidas2011}

\bibitem[{{Vourlidas} et~al.(2013){Vourlidas}, {Lynch}, {Howard}, and
  {Li}}]{vourlidas2013}
{Vourlidas}, A., {Lynch}, B.~J., {Howard}, R.~A., and {Li}, Y. (2013).
\newblock {How Many CMEs Have Flux Ropes? Deciphering the Signatures of Shocks,
  Flux Ropes, and Prominences in Coronagraph Observations of CMEs}.
\newblock \emph{\solphys} 284, 179--201.
\newblock \doi{10.1007/s11207-012-0084-8}
\bibAnnoteFile{vourlidas2013}

\bibitem[{{Vourlidas} and {Webb}(2018)}]{vourlidas2018}
{Vourlidas}, A. and {Webb}, D.~F. (2018).
\newblock {Streamer-blowout Coronal Mass Ejections: Their Properties and
  Relation to the Coronal Magnetic Field Structure}.
\newblock \emph{\apj} 861, 103.
\newblock \doi{10.3847/1538-4357/aaca3e}
\bibAnnoteFile{vourlidas2018}

\bibitem[{{Yurchyshyn} et~al.(2009){Yurchyshyn}, {Abramenko}, and
  {Tripathi}}]{yurchyshyn2009}
{Yurchyshyn}, V., {Abramenko}, V., and {Tripathi}, D. (2009).
\newblock {Rotation of White-light Coronal Mass Ejection Structures as Inferred
  from LASCO Coronagraph}.
\newblock \emph{\apj} 705, 426--435.
\newblock \doi{10.1088/0004-637X/705/1/426}
\bibAnnoteFile{yurchyshyn2009}

\bibitem[{{Zhang} et~al.(2007){Zhang}, {Richardson}, {Webb}, {Gopalswamy},
  {Huttunen}, {Kasper} et~al.}]{zhang2007}
{Zhang}, J., {Richardson}, I.~G., {Webb}, D.~F., {Gopalswamy}, N., {Huttunen},
  E., {Kasper}, J.~C., et~al. (2007).
\newblock {Solar and interplanetary sources of major geomagnetic storms (Dst
  $\leq -100$ nT) during 1996-2005}.
\newblock \emph{\jgr} 112, A10102.
\newblock \doi{10.1029/2007JA012321}
\bibAnnoteFile{zhang2007}

\bibitem[{{Zhukov} and {Auch{\`e}re}(2004)}]{zhukov2004}
{Zhukov}, A.~N. and {Auch{\`e}re}, F. (2004).
\newblock {On the nature of EIT waves, EUV dimmings and their link to CMEs}.
\newblock \emph{\aap} 427, 705--716.
\newblock \doi{10.1051/0004-6361:20040351}
\bibAnnoteFile{zhukov2004}

\end{thebibliography}

\end{document}